\documentclass[11pt]{article}
\usepackage{amsfonts}

\usepackage{geometry}
\usepackage{mathrsfs}
\usepackage{amsmath}
\usepackage{graphicx}
\usepackage{makecell}
\usepackage{fancyhdr}
\usepackage{algorithm}
\usepackage{algpseudocode}
\usepackage{booktabs}
\usepackage{bbm}
\usepackage{CJK}
\usepackage{cite}
\usepackage[round, sort]{natbib}

\newcommand{\eqnb}{\begin{equation}}
	\newcommand{\eqne}{\end{equation}}

\newtheorem{The}{Theorem}

\newtheorem{Lem}{Lemma}

\begin{document}
	
	\title{A Closed Queueing Maintenance Network with Two Batch	Policies}
	\author{Rui-Na Fan \\
		School of Management\\
		Fudan University, Shanghai 200433, China\\
		Quan-Lin Li\\
		School of Economics and Management\\
		Beijing University of Technology, Beijing 100124, China\\
		Xiaole Wu\\
		School of Management\\
		Fudan University, Shanghai 200433, China\\
		Zhe George Zhang*\\
		Department of Decision Sciences, Western Washington University\\
		Bellingham, Washington 98225, USA\\
		and\\
		Beedie School of Business, Simon Fraser University\\
		Burnaby, British Columbia V5A 1S6, Canada}
	\maketitle

\begin{abstract}
This paper discusses a maintenance network with failed items
that can be removed, repaired, redistributed, and reused under two batch
policies: one for removing the failed items from each base to a maintenance
shop and the other for redistributing the repaired items from the
maintenance shop to bases. This maintenance network can be considered a virtual closed queueing network, 
and the
Markov system of each node is described as an elegant block-structured
Markov process whose stationary probabilities can be computed by the
RG-factorizations. The structure of this maintenance network is novel and interesting. To compute the closed queueing 
network,
we set
up a new nonlinear matrix equation to determine the relative arrival rates, in which the nonlinearity comes from two different 
groups of
processes: the failure and removal processes and the repair and
redistribution processes. This paper also extends a simple queueing system of a node to a more general
block-structured Markov process which can be computed by the RG-factorizations. Based on this, the paper establishes a more 
general
product-form solution for the closed queueing network and provides
performance analysis of the maintenance network. Our method will open a new
avenue for quantitative evaluation of more general maintenance networks.
\end{abstract}



\section{Introduction}

 With the development of the sharing economy, various sharing systems
are prevalent; for example, there are bike-sharing systems, car-sharing systems, shared
power banks, umbrella-sharing systems, shared sleep warehouses, etc. These sharing
systems provide customers with convenience and value of service. However, with the operation of
these systems, maintenance problems become critical for the
sustainable development. Since stations or sites are
scattered throughout a city, the removal, repair, and redistribution
processes are labor-consuming.  Finding the efficient removing and reusing process is vital to these sharing systems. 

To perform a quantitative analysis, this paper
develops a closed queueing maintenance network from these sharing systems. It is found that the structure of the closed queueing 
network is novel and interesting. Due to the complicated and special structure, the modeling and analysis of the maintenance 
network is quite difficult.  To the best of authors' knowledge, there are no existing methodologies or results along
such a research line. We consider two reasonable and practical batch policies, which make the removing and redistributing more 
efficient, although it introduces the complexity to the model. 

For such a maintenance network, there are some
interesting practical problems that need to be addressed. First, how to
estimate the effect of item failure on the quality of service and on
the system performance. Second, how to develop effective removal policies of failed
items such that the system performance can further be improved. To answer these questions, this paper 
derives the product-form
solution of the closed queueing network by solving the non-linear matrix equation system for the relative arrival rates combined 
with the RG-factorizations of block-structured Markov processes. Based on the product-form solution, the performance measures for 
the maintenance network can be computed.

Comparing with the existing studies in the literature of maintenance networks, our analysis
has the following four key features.

\textit{A novel maintenance network: }For the maintenance
network of sharing systems, items may fail at any base, failed
items are batch-removed from bases to the maintenance shop, and
repaired items are batch-redistributed for reuse from the maintenance
shop to bases. So far, there have been no studies of using analytical models on such a complicated maintenance network. 

\textit{A new class of virtual closed queueing networks:} Compared with the previous studies on the closed queueing network 
\cite{li2016unified, li2017nonlinear,
	li2017fluid}. The features of the items' failure, removal,
repair, distribution, and reuse processes, together with two batch policies,
can
substantially change the physical structure of the virtual closed queueing
network. This paper is the first to analyze closed queueing networks with such an interesting and complicated structure.

\textit{The block-structured Markov process: }This paper also contributes to
the literature of closed queueing networks by extending and
generalizing a simple queueing system (e.g., the M/M/1 queue, the M/M/C
queue, the M/G/1 queue, and others) of a node to a more general
block-structured Markov process. We show that such a block structure is
established by either items' failure and batch removal processes in a
base or items' repair and batch redistribution processes in the
maintenance shop. 

\textit{A nonlinear routing matrix equation:} In the theory of closed
queueing networks, it is well known that the relative arrival rates can
uniquely be determined by a system of linear equations $\mathbf{e}P=\mathbf{e%
}$, where $\mathbf{e}$ is the relative arrival rate vector and $P$ is the
routing matrix, e.g., see \cite{bolch2006queueing} and \cite%
{serfozo2012introduction} for details. However, from the
maintenance network of sharing systems, we find a new
fundamental result: The relative arrival rate vector should be determined by a
nonlinear routing matrix equation $\mathbf{e}P\left( \mathbf{e}\right) =%
\mathbf{e}$, where some entries of the routing matrix $P\left( \mathbf{e}%
\right) $ depend on the relative arrival rate vector. Therefore, the nonlinearity of the routing matrix
equation opens a new research avenue in the study of closed queueing
networks.

This paper extends and generalizes the
product-form solution of the closed queueing networks to a more general case
that the Markov systems of the nodes are described as the block-structured
Markov processes.
The methodology and results given in this paper shed
light on the study of more general maintenance networks
and open a new research direction.

The remainder of this paper is organized as follows. Section 2 provides a
literature review. Section 3 presents a model description for a maintenance
network with failed items. Section 4 formulates the maintenance network as a
closed queueing network, and provides a detailed analysis for the relative
arrival rates, the service processes, the block-structured Markov processes,
and the nonlinear routing matrix equation. Section 5 derives the
product-form solution of the closed queueing network which is used for performance
analysis of the maintenance network. 
Section 6 concludes the paper with a summary.

\section{Literature Review}

\label{literature-review}

Our current work is related to three streams of literature:  maintenance networks and reliability analysis, closed queueing 
networks, and the
RG-factorizations. We review some related literature in these areas.

\textbf{Maintenance networks and reliability analysis: }\cite{heidergott2010gradient} studied the gradient estimation for 
multicomponent maintenance systems with age-replacement 
policy. \cite{you2019generalized} provided a maintenance policy for 
maintenance scheduling with the help of both time-based maintenance and condition-based maintenance techniques. 
 \cite{rawat2020joint, rawat2020simulation} presented a joint optimization 
approach 
of reliability design and level of repair 
analysis for fleet systems with multi-machine and multi-indenture.
\cite{lin2020pso} studied the optimal maintenance plan by examining the relationship
between facility reliability and lifespan. There has been much research on manufacturing engineering with
maintenance and reliability, such as manufacturing systems with
repairable machines by \cite{buzacott1986queueing} and \cite{gregory2015manufacturing}, production systems
with repairable bases by \cite{li2008production}, and parts inventory
systems with repairable parts by \cite{park2011multi, park2014approximation}. In contrast to the literature, this paper studies a 
maintenance network 
with a special and novel structure.

\textbf{Closed queueing networks:} 
 \cite{george2011fleet} modeled the vehicle rental systems as a closed queueing network to  determine the optimal number of
 parking places in each rental location. \cite{waserhole2016pricing,waserhole2013pricing} used closed queueing networks, 
 combined with the fluid approximation, to establish Markov
decision models to determine the optimal policy of the bike-sharing system.  As a closely related application of closed queueing 
networks, the sharing systems have drawn much research attention. Important
examples include \cite{adelman2007price}, \cite{george2012stochastic}, \cite{fanti2014fleet}, \cite{samet2018performance},
and so forth. Also, closed queueing networks are used in the research of maintenance systems.  
\cite{park2011multi,park2014approximation} 
established a multi-class closed queueing maintenance network model with a parts
inventory system. \cite{gross1983closed} and \cite{madu1988closed} used a closed queueing network to decide the number of 
repairable items and the capacity of repair depot. This paper models a maintenance network as a closed queueing network and 
extends some nodes from simple queueing
system to general block-structured Markov processes.

\textbf{RG-factorizations:} For block-structured Markov processes, \cite%
{li2010constructive} provided a unified effective computational framework of
the RG-factorizations, which includes stationary performance analysis,
transient solution, the first passage times, reward processes,
quasi-stationary distribution, and so forth. For further details on
RG-factorization and their usefulness in stochastic modeling, readers can refer
to \cite{wang2007queueing}, \cite{li2009performance}, \cite{li2005light}, \cite{wang2010per}, 
\cite{yu2015algorithm}, \cite{samanta2021analysis}, and \cite{das2021modelling}.

This paper studies a maintenance system with a novel structure by applying the RG-factorizations of
block-structured Markov processes to the closed queueing networks and contributes to the literature by providing a more
general product-form solution.

\section{Model Description}

In this section, we describe a maintenance network with $N$ bases and one
maintenance shop. The $N$ bases are assumed to be different and the capacity of
each base is sufficiently large. Items are kept and may fail in the bases. The total number of items in the
maintenance network is fixed at $K$. See Figure 1 for a pictorial
illustration.

\begin{figure}[h]
	\centering                         \includegraphics[width=9cm]{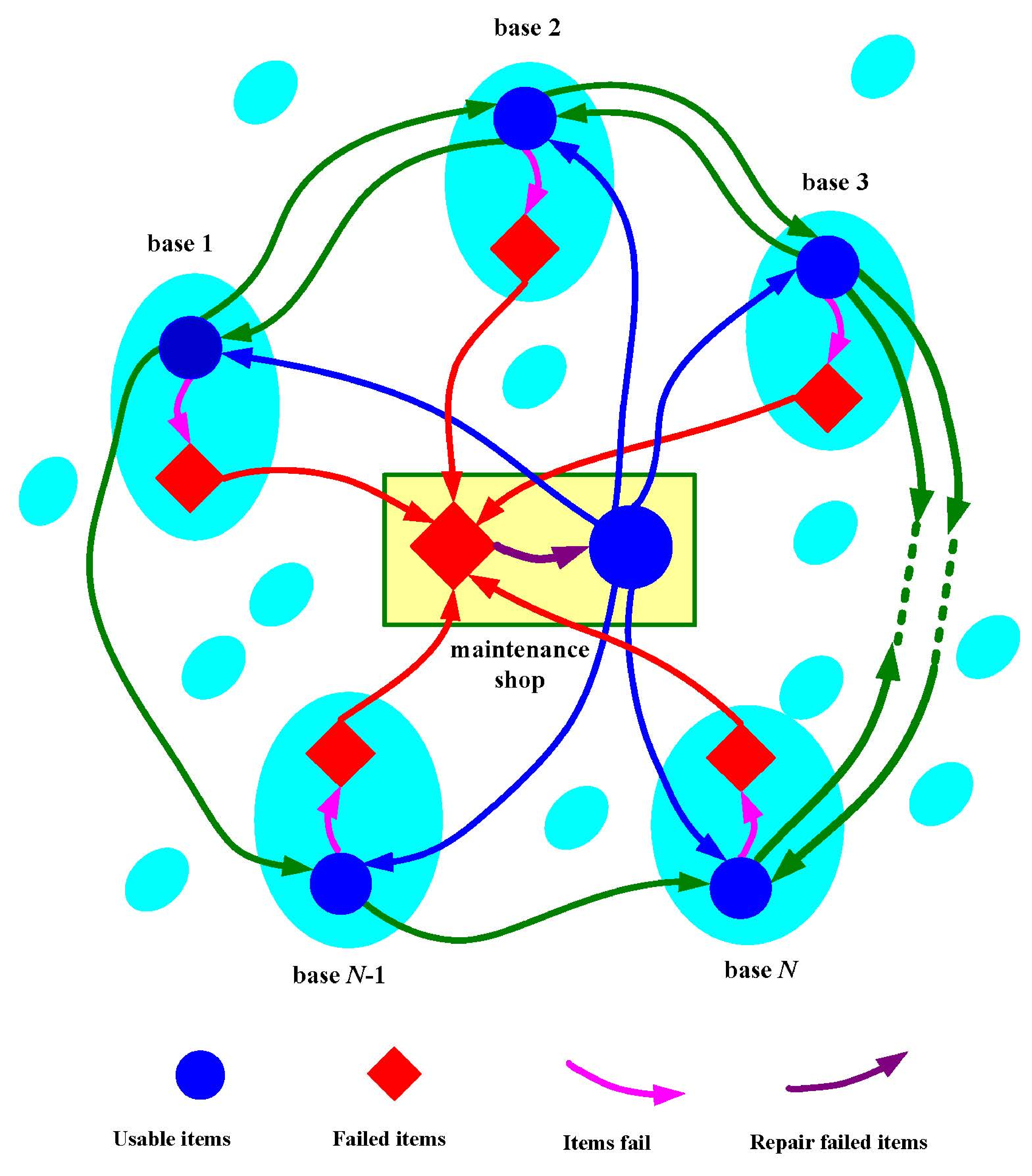} 
	\newline
			\setlength{\abovecaptionskip}{-0.3cm} 
	\setlength{\belowcaptionskip}{-0.3cm} 
	\caption{The physical structure of the maintenance network}
	\label{figure:figure-1}
\end{figure}

We denote base $i$ as Node $i$, $i=1,\ldots,N$ and express the maintenance shop as Node $0$.
There are transfer path nodes between any two nodes, which can be divided into the following two kinds.

\textit{(i) Path nodes between bases.} Let Node $i\rightarrow j$ denote the
path node from base $i$ to base $j$. Note that Node $i\rightarrow j$ and Node
$j\rightarrow i$ may be different due to practical factors.

Denote all the nodes beginning from base $i$ for $1\leq i\leq N$ as
$R_{B}\left(  i\right)  =\{$Node $i\rightarrow j:j\neq i,1\leq j\leq N\}$. Write all the bases in the downlink of base $i$ as 
$\Theta_{i}%
=\{$Node $j:$ Node $i\rightarrow j\in R_{B}\left(  i\right)  \}$.

\textit{(ii) Path nodes between bases and the maintenance shop.} There are two classes of path nodes: (ii-1) the path nodes for
removing failed items from a base to the maintenance shop, denoted as
$R_{E}\left(  0\right)  =\{$Node $i\rightarrow0:1\leq i\leq N\}$; (ii-2) and the
path nodes for redistributing repaired items from the maintenance shop to
bases, written as $R_{B}\left(  0\right)  =\{$Node $0\rightarrow i:1\leq i\leq
N\}$. 

An outside user arrives at base $i$ in order to rent an item. If there is no
usable item (either the base $i$ is empty or all the items in base $i$ are
failed), then the user immediately leaves the system. If there is at least
one usable item in base $i$, then the user rents an item and gets the
service on Node $i\rightarrow j$ with probability $p_{i,j}$ for $j\neq i$, $%
1\leq j\leq N$ and $\sum\nolimits_{j\in \Theta _{i}}p_{i,j}=1$ for each $%
i=1,2,\ldots ,N$. The arrivals of outside users at base $i$\ follow a
Poisson process with arrival rate $\lambda _{i}>0$ for $1\leq i\leq N$. We
assume that the service times on Node $i\rightarrow j$ are i.i.d. and
exponential with rate $\mu _{i,j}>0$. We assume that all the items are
identical, and the lifetime of an item is exponential with failure rate $%
\alpha >0$. For the maintenance problem, we propose the following two batch
policies.

\textbf{A batch removal policy:} Once the number of failed items at any
base reaches a positive integer $M$, the $M$
failed items are removed in a batch and transported to the maintenance shop. We
assume the transportation time on Node $i\rightarrow 0$ is
exponentially distributed with rate $\mu _{i,0}>0$.

\textbf{A batch redistribution policy: }Once the number of repaired items in
the maintenance shop reaches a given positive integer $Z$, the $Z$ repaired
items are taken in a batch away from the maintenance shop to some bases, in
which $Z_{i}$ repaired items are sent to the base $i$ with $%
\sum_{i=1}^{N}Z_{i}=Z$. Let $\beta _{i}=Z_{i}/Z$. The transportation time from the maintenance shop to base $i$ is exponentially 
distributed
with rate $\mu
_{0,i}>0$.

If a seriously damaged item is scrapped after a repair, then a new item
is added to the system immediately. Thus, the total number of items in the system is
constant. We assume that there are $r$ repairmen in the maintenance shop, and
the repair time of each failed item is exponentially distributed with rate $w>0$. For
convenience of expression, we assume that $Z=\psi M$, where $\psi $ is a
given positive integer. In addition, let $\phi =\lfloor K/M\rfloor $. 

We assume that all the random variables mentioned above are independent of
each other. The notation is summarized in Table 1.

\begin{table}[h!]
	\begin{center}
  \caption{Summary of Notation}
	\begin{tabular}{ll}		
				\hline
		$N	$&   Number of bases;\\
		$K $	&   Total number of items;\\
		$\Theta _{i}$	& Set of bases in the downlink of base $i$, for $1\leq i\leq N$;\\
		$\lambda _{i}$	& User arrival rate at base $i$, for $1\leq i\leq N$; \\
		$\mu _{i,j} $	&  Service rate on path node $i\rightarrow j$, for $1\leq i\leq N, j\in\Theta_{i}$;\\
			$\mu _{i,0}$	&  Service rate on path node $i\rightarrow 0$, for $1\leq i\leq N$;\\
		$	\mu _{0,i} $	&Service rate on path node $0\rightarrow i$, for $1\leq i\leq N$;  \\
		$p_{i,j}$	& Transition probability from base $i$ to path node $i\rightarrow j $, for $1\leq i\leq N, j\in\Theta_{i}$;\\
		$	\alpha $	&Failure rate of an item; \\
		$w$	&  Repair rate of an failed item;\\
		$r $	& Number of repairmen in the maintenance shop; \\
		$M $	&Batch size of removal failed items from any base;  \\
		$Z $	& Batch size of redistributing repaired items from the maintenance shop; \\
		$Z_{i}$	& Number of repaired items redistributed to base $i$, for $1\leq i\leq N$;\\
		$\beta _{i}$	&  $ Z_{i}/Z $;\\
		$\phi $	&$ \lfloor K/M\rfloor$  \\ 
		$	\psi$	&$Z/M$, a positive integer.  \\ \hline
	\end{tabular}
\end{center}
\end{table}
\section{A Virtual Closed Queueing Network}

In this section, we describe the maintenance network as a
closed queueing network, explain its physical structure, and introduce mathematical
notations.

Since the total number of items in the maintenance
network is fixed, we can formulate the system as a closed queueing network as
follows.

\textbf{(1) Virtual customers:} Since items are either kept in bases, transferred on path nodes, or
repaired in the maintenance shop, they can be viewed as virtual customers. The items have two states: usable and
failed. We use $G$ ($G$: Good) and $B$ ($B$: Bad) to denote the usable
and failed items, respectively.

\textbf{(2) Virtual nodes:} Note that bases, path nodes, and the maintenance
shop have different physical attributes such as functions and geographical
structures. Thus they are considered as different classes of nodes.

\textbf{(3) An irreducible path graph:} The set of all the virtual nodes in the
maintenance network is given by $\Theta =\{ \text{Node }i:0\leq i\leq N\} \underset{i=1}{\overset{N%
		}{\cup }}R_{B}\left( i\right) \cup R_{B}\left( 0\right) \cup R_{E}\left(
	0\right)$. We assume that all the path nodes of the system are connected as an
irreducible path graph whose nodes are in the set $\Theta $. In this case,
we call the maintenance network path irreducible.

Let $Q_{G}^{\left( i\right) }\left( t\right) $ and $Q_{B}^{\left( i\right)
}\left( t\right) $ denote the numbers of usable items and failed items
kept in base $i$ at time $t\geq 0$ for $1\leq i\leq N$, respectively; and $%
Q_{G}^{\left( 0\right) }\left( t\right) $ and $Q_{B}^{\left( 0\right)
}\left( t\right) $ the numbers of usable items and failed items in the
maintenance shop at time $t\geq 0$, respectively. Let $R_{i,j}\left(
t\right) $ be the number of items transfered on Node $i\rightarrow j$ at
time $t$ for $1\leq i\leq N$ and $j\in \Theta _{i}$. $R_{i,0}\left( t\right) 
$ and $R_{0,i}\left( t\right) $ are numbers of items on Node $i\rightarrow 0$
and Node $0\rightarrow i$\ at time $t$ for $1\leq i\leq N$, respectively.

Denote the state vector by $X\left(  t\right)  =(\mathbf{L}_{0}\left(  t\right)  ,\mathbf{L}%
_{1}\left(  t\right)  ,\mathbf{L}_{2}\left(  t\right)  ,\ldots,\mathbf{L}%
_{N-1}\left(  t\right)  ,\mathbf{L}_{N}\left(  t\right)  )$ , where
$\mathbf{L}_{0}\left(  t\right)  =(Q_{G}^{\left(  0\right)  }\left(  t\right)
,\\
Q_{B}^{\left(  0\right)  }\left(  t\right)  ;R_{i,0}\left(  t\right)
,R_{0,i}\left(  t\right)  :1\leq i\leq N)$, and for $1\leq i\leq N$,
$\mathbf{L}_{i}\left(  t\right)  =(Q_{G}^{\left(  i\right)  }\left(  t\right)
,Q_{B}^{\left(  i\right)  }\left(  t\right)  ,R_{i,j}\left(  t\right)
:j\in\Theta_{i})$.
Obviously, $\left\{ X\left( t\right) :t\geq 0\right\} $ is a continuous-time
irreducible Markov process. The state space of the Markov process $\left\{
X\left( t\right) :t\geq 0\right\} $ is given by
\begin{align*}
	\setlength{\abovedisplayskip}{3pt}
	\Omega & =\left\{  \mathbf{n}:\sum_{i=0}^{N}\left(  n_{G}^{\left(  i\right)
	}+n_{B}^{\left(  i\right)  }\right)  +\sum_{i=1}^{N}\sum_{j\in\Theta_{i}%
	}m_{i,j}+\sum_{i=1}^{N}\left(  m_{i,0}+m_{0,i}\right)  =K,\right.  \\
	& \left.  0\leq n_{B}^{\left(  i\right)  }\leq M,0\leq n_{G}^{\left(
		i\right)  },m_{i,j}\leq K,0\leq n_{G}^{\left(  0\right)  }\leq Z,0\leq
	n_{B}^{\left(  0\right)  }\leq\phi M,\right.  \\
	& \left.  n_{G}^{\left(  0\right)  }+n_{B}^{\left(  0\right)  }=kM,m_{i,0}%
	=kM,m_{0,i}=lZ_{i},\text{for }1\leq i\leq N,j\in\Theta_{i},\right.  \\
	& \left.  0\leq k\leq\phi,0\leq l\leq\phi/\psi,0\leq Z_{i}\leq Z\right\}  ,
	\setlength{\belowdisplayskip}{3pt}
\end{align*}
where $\mathbf{n}=(\mathbf{n}_{0},\mathbf{n}_{1},\mathbf{n}_{2},\ldots
,\mathbf{n}_{N-1},\mathbf{n}_{N})$, and $\mathbf{n}_{0}=(n_{G}^{\left(
	0\right)  },n_{B}^{\left(  0\right)  };m_{i,0},m_{0,i}:1\leq i\leq N)$, for
$1\leq i\leq N$, $\mathbf{n}_{i}=(n_{G}^{\left(  i\right)  },n_{B}^{\left(
	i\right)  };m_{i,j}:j\in\Theta_{i})$.

Note that $n_{G}^{\left( i\right) }$ and $n_{B}^{\left( i\right) }$\ are the
numbers of usable items and failed items kept in base $i$ for $1\leq
i\leq N$, respectively; $n_{G}^{\left( 0\right) }$ and $n_{B}^{\left(
	0\right) }$ are the numbers of usable items and failed items in the
maintenance shop, respectively; $m_{i,j}$ is the number of items ridden on
Node $i\rightarrow j$ for $1\leq i\leq N$, $j\in \Theta _{i}$; and $%
m_{i,0},m_{0,i}$ denote the numbers of failed items and repaired items
transported on Node $i\rightarrow 0$ and Node $0\rightarrow i$ for $1\leq
i\leq N$, respectively. Figure 2 illustrates the physical structure of the closed queueing network.

To study the closed queueing network, we analyze the \textit{relative arrival rates} in Subsection 4.1, the Markov processes of any 
base in 
Subsection 4.2,  the Markov processes of the maintenance shop in Subsection 4.3, and the routing matrix in Subsection 4.4.
\begin{figure}[h]
	\centering                         \includegraphics[width=8cm]{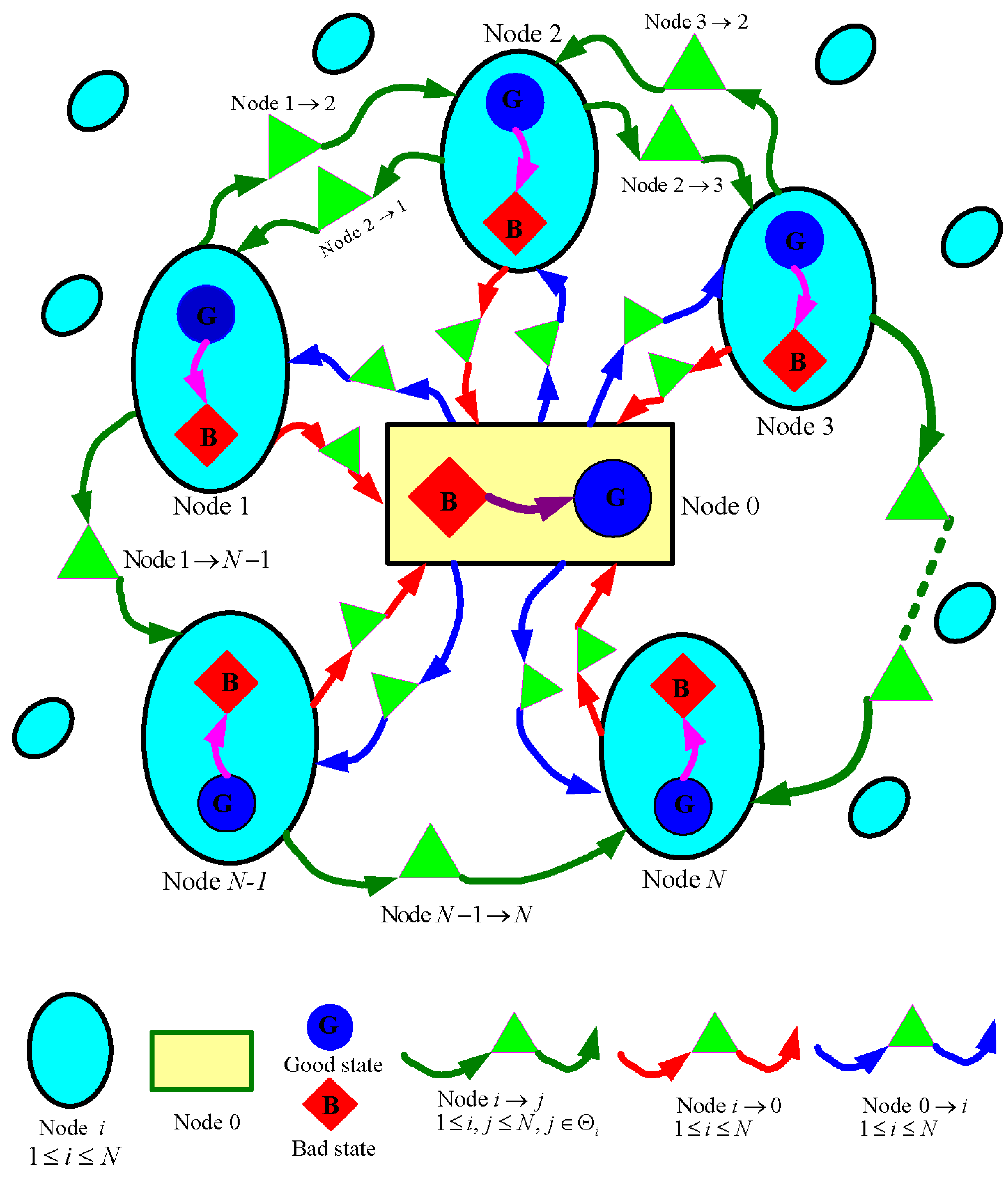} 
	\newline
		\setlength{\abovecaptionskip}{-0.3cm} 
\setlength{\belowcaptionskip}{-0.3cm} 
	\caption{The structure of the virtual closed queueing network}
	\label{figure:figure-2}
\end{figure}
\subsection{The relative arrival rates}

\label{the-virtual}

Note that the relative arrival rates play a key role in deriving
the product-form solution of queueing networks, e.g., see \cite%
{bolch2006queueing} and \cite{serfozo2012introduction} for details.

We denote by $e_{i}$ the relative arrival rate of Node $i$ for $0\leq i\leq
N $, $e_{i,j}$ the relative arrival rate of Node $i\rightarrow j$ for $1\leq
i\leq N$ and $j\in \Theta _{i}$, $e_{i,0}$ and $e_{0,i}$ the relative
arrival rates of Node $i\rightarrow 0$ and Node $0\rightarrow i$,
respectively. Let%
\begin{equation*}
		\setlength{\abovedisplayskip}{3pt}
	\mathbb{E=}\left\{ \mathbf{e}_{0},\mathbf{e}_{1},\mathbf{e}_{2},\ldots ,%
	\mathbf{e}_{N-1},\mathbf{e}_{N}\right\} ,
	\setlength{\belowdisplayskip}{3pt}
\end{equation*}%
where $\mathbf{e}_{0}=(e_{0};e_{i,0},e_{0,i}:1\leq i\leq N)$, and for $1\leq
i\leq N$, $\mathbf{e}_{i}=(e_{i};e_{i,j}:$ $j\in\Theta_{i})$. 

Note that the relative arrival rates only depend on the physical structure
of the closed queueing network, while they are independent of the states of
the Markov process $\left\{ X\left( t\right) :t\geq 0\right\} $.

To determine the relative arrival rates, we first analyze the Markov
processes of various bases and of the maintenance shop in Subsections 4.2
and 4.3, respectively. Then we establish the routing matrix in
Subsection 4.4, so that we can set up a nonlinear routing matrix equation to
determine the relative arrival rates.

\subsection{Markov process of any base}

\label{markov-parking}

When the relative arrival rates are introduced to various nodes in the
closed queueing network, all the nodes are isolated from each other so that
the Markov processes of the nodes are independent of each other. Therefore,
the Markov system of each node is a block-structured Markov process, which
is based on the numbers of usable items and failed items in either any
base or the maintenance shop.

To set up the block-structured Markov process of a base, it is necessary to analyze
the state transitions of the Markov process for the base. Note that $%
Q_{G}^{\left( i\right) }\left( t\right) $ and $Q_{B}^{\left( i\right)
}\left( t\right) $ are the numbers of usable items and failed items at
base $i$ at time $t$, respectively. Under the batch removal policy of
failed items with parameter $M$, it is easy to see that $\left\{
\left( Q_{G}^{\left( i\right) }\left( t\right) ,Q_{B}^{\left( i\right)
}\left( t\right) \right) :t\geq 0\right\} $ is a two-dimensional Markov
process with finite levels and phases, where $Q_{G}^{\left( i\right) }\left(
t\right) $ is the phase variable while $Q_{B}^{\left( i\right) }\left(
t\right) $ is the level variable. Figure 3 depicts the state transition
relations of the Markov process. Thus the infinitesimal
generator $\mathbf{Q}$ of the Markov process is given by
\begin{equation}
		\setlength{\abovedisplayskip}{3pt}
	\mathbf{Q}=\left( 
	\begin{array}{ccccc}
		Q_{0,0} & Q_{0,1} &  &  &  \\ 
		& Q_{1,1} & Q_{1,2} &  &  \\ 
		&  & \ddots & \ddots &  \\ 
		&  &  & Q_{M-1,M-1} & Q_{M-1,M} \\ 
		Q_{M,0} &  &  &  & Q_{M,M}%
	\end{array}%
	\right) .  \label{V-1}
	\setlength{\belowdisplayskip}{3pt}
\end{equation}%
All the elements of the infinitesimal generator $\mathbf{Q}$ are
given in Appendix A.
\begin{figure}[h]
	\centering                         \includegraphics[width=14cm]{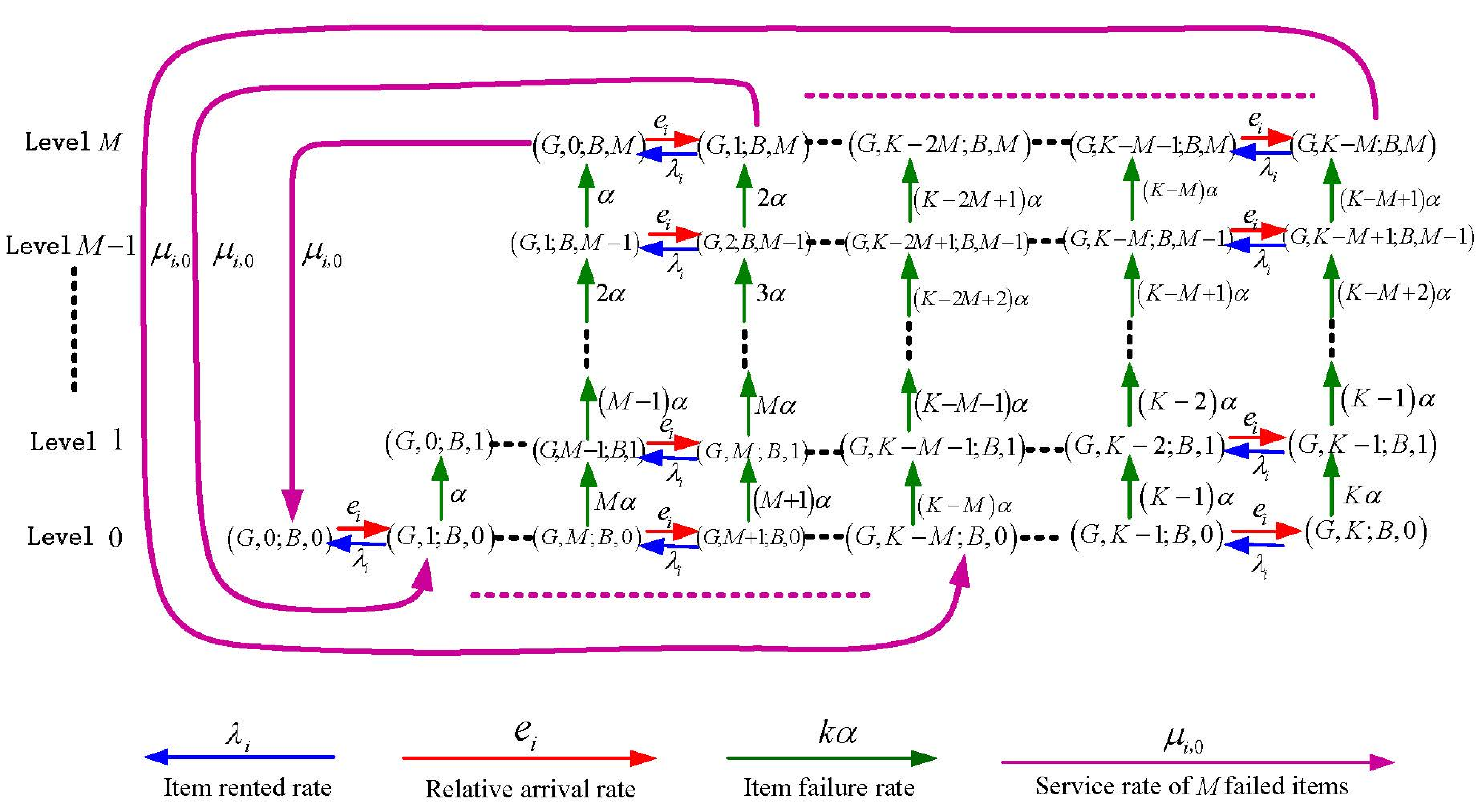} 
	\newline
	\setlength{\abovecaptionskip}{-0.3cm} 
	\setlength{\belowcaptionskip}{-0.3cm} 
	\caption{State transition relations of the Markov process in base $i$ }
	\label{figure:figure-3}
\end{figure}

The structure of the matrix $\mathbf{Q}$ is bidiagonal blocks
with a special block $Q_{M,0}$ in the lower-left corner. Obviously, the
matrix-geometric method (see \cite{neuts1994matrix} or \cite%
{latouche1999introduction}) does not apply to such a Markov process. Thus, to compute the stationary probabilities of the 
Markov
process, we need to use the UL-type RG-factorization (see \cite%
{li2010constructive} for details). To this end, we write $	Q_{M,M}^{[\leq M]}=Q_{M,M}$,
and for $1\leq k\leq M-1$, 

\begin{equation*}
	Q^{[\leq k]}=\left( 
	\begin{array}{ccccc}
		Q_{0,0} & Q_{0,1} &  &  &  \\ 
		& Q_{1,1} & Q_{1,2} &  &  \\ 
		&  & \ddots & \ddots &  \\ 
		&  &  & Q_{k-1,k-1} & Q_{k-1,k} \\ 
		Q_{k,0}^{[\leq k]} &  &  &  & Q_{k,k}%
	\end{array}%
	\right) .
\end{equation*}

\begin{Lem}
	We have $Q_{M,M}^{[\leq M]}=Q_{M,M}$, for $1\leq k\leq M-1$,
	\begin{equation}
			\setlength{\abovedisplayskip}{3pt}
		Q_{k,0}^{[\leq k]}=\prod\nolimits_{l=k}^{M-1}Q_{l,l+1}\left(
		-Q_{l+1,l+1}\right) ^{-1}Q_{M,0}  \label{eq-Qk}
		\setlength{\belowdisplayskip}{3pt}
	\end{equation}
	and 
	\begin{equation}
			\setlength{\abovedisplayskip}{3pt}
		Q^{\left[ \leq 0\right] }=Q_{0,0}+Q_{0,1}\left( -Q_{1,1}\right) ^{-1}\left[
		\prod\limits_{l=1}^{M-1}Q_{l,l+1}\left( -Q_{l+1,l+1}\right) ^{-1}\right]
		Q_{M,0}.  \label{eq-Q0}
		\setlength{\belowdisplayskip}{3pt}
	\end{equation}
\end{Lem}

The proof is given in Appendix B.

Let%
\begin{equation}
		\setlength{\abovedisplayskip}{3pt}
	\Psi _{n}=Q_{n,n}^{[\leq n]},\text{ }0\leq n\leq M,  \label{eq-RG}
	\setlength{\belowdisplayskip}{3pt}
\end{equation}%
\begin{equation}
		\setlength{\abovedisplayskip}{3pt}
	R_{i,j}=Q_{i,j}^{\left[ \leq j\right] }\left( -\Psi _{j}\right) ^{-1},\text{ 
	}0\leq i<j\leq M  \label{eq-R}
\setlength{\belowdisplayskip}{3pt}
\end{equation}%
and%
\begin{equation}
		\setlength{\abovedisplayskip}{3pt}
	G_{i,j}=\left( -\Psi _{i}\right) ^{-1}Q_{i,j}^{\left[ \leq i\right] },\text{ 
	}0\leq j<i\leq M.  \label{eq-G}
\setlength{\belowdisplayskip}{3pt}
\end{equation}

The following theorem gives the UL-type RG-factorization of the Markov
process $\mathbf{Q}$. It is easy to see that the RG-factorization has a
nice block structure.

\begin{The}
	The UL-type RG-factorization of the continuous-time Markov process $\mathbf{%
		Q }$ is given by%
	\begin{equation*}
			\setlength{\abovedisplayskip}{3pt}
		\mathbf{Q}=\left( I-R_{U}\right) \Psi _{D}\left( I-G_{L}\right) ,
		\setlength{\belowdisplayskip}{3pt}
	\end{equation*}
	where%
	\begin{equation*}
			\setlength{\abovedisplayskip}{3pt}
		R_{U}=\left( 
		\begin{array}{ccccccc}
			0 & R_{0,1} &  &  &  &  &  \\ 
			& 0 & R_{1,2} &  &  &  &  \\ 
			&  & 0 & R_{2,3} &  &  &  \\ 
			&  &  & \ddots & \ddots &  &  \\ 
			&  &  &  & 0 & R_{M-2,M-1} &  \\ 
			&  &  &  &  & 0 & R_{M-1,M} \\ 
			&  &  &  &  &  & 0%
		\end{array}
		\right) ,
		\setlength{\belowdisplayskip}{3pt}
	\end{equation*}%
	\begin{equation*}
			\setlength{\abovedisplayskip}{3pt}
		\Psi _{D}=\text{diag}\left( \Psi _{0},\Psi _{1},\Psi _{2},\ldots ,\Psi
		_{M-1},\Psi _{M}\right) ,
		\setlength{\belowdisplayskip}{3pt}
	\end{equation*}
	and%
	\begin{equation*}
			\setlength{\abovedisplayskip}{3pt}
		G_{L}=\left( 
		\begin{array}{ccccccc}
			0 &  &  &  &  &  &  \\ 
			G_{1,0} & 0 &  &  &  &  &  \\ 
			G_{2,0} &  & 0 &  &  &  &  \\ 
			\vdots &  &  & \ddots &  &  &  \\ 
			G_{M-2,0} &  &  &  & 0 &  &  \\ 
			G_{M-1,0} &  &  &  &  & 0 &  \\ 
			G_{M,0} & 0 & 0 & \cdots & 0 & 0 & 0%
		\end{array}
		\right) .
		\setlength{\belowdisplayskip}{3pt}
	\end{equation*}
\end{The}

The proof is provided in Appendix C.

Since the Markov process is irreducible, with a finite state space, and satisfies $%
\mathbf{Q}\mathbbm{1}=0$ where $\mathbbm{1}$ is a column vector of ones, it
must be positive recurrent. Let $\mathbf{\pi }_{B}^{\left( i\right) }=\left( \mathbf{\pi }_{B,0}^{\left(
	i\right) },\mathbf{\pi }_{B,1}^{\left( i\right) },\mathbf{\pi }%
_{B,2}^{\left( i\right) },\right. \\\left. \ldots ,\mathbf{\pi }%
_{B,M-1}^{\left( i\right) },\mathbf{\pi }_{B,M}^{\left( i\right) }\right) $
be the stationary probability vector of the Markov process,
where%
\begin{equation*}
		\setlength{\abovedisplayskip}{3pt}
	\left\{ 
	\begin{array}{l}
		\mathbf{\pi }_{B,0}^{\left( i\right) }=\left( \pi _{G,0;B,0}^{\left(
			i\right) },\pi _{G,1;B,0}^{\left( i\right) },\pi _{G,2;B,0}^{\left( i\right)
		},\ldots ,\pi _{G,K-1;B,0}^{\left( i\right) },\pi _{G,K;B,0}^{\left(
			i\right) }\right) , \\ 
		\mathbf{\pi }_{B,1}^{\left( i\right) }=\left( \pi _{G,0;B,1}^{\left(
			i\right) },\pi _{G,1;B,1}^{\left( i\right) },\pi _{G,2;B,1}^{\left( i\right)
		},\ldots ,\pi _{G,K-2;B,1}^{\left( i\right) },\pi _{G,K-1;B,1}^{\left(
			i\right) }\right) , \\ 
		\text{ \ \ \ \ \ \ \ \ \ \ \ \ \ \ \ \ \ \ \ \ \ \ \ \ \ }\vdots \\ 
		\mathbf{\pi }_{B,M}^{\left( i\right) }=\left( \pi _{G,0;B,M}^{\left(
			i\right) },\pi _{G,1;B,M}^{\left( i\right) },\pi _{G,2;B,M}^{\left( i\right)
		},\ldots ,\pi _{G,K-M-1;B,M}^{\left( i\right) },\pi _{G,K-M;B,M}^{\left(
			i\right) }\right) .%
	\end{array}%
	\right.
	\setlength{\belowdisplayskip}{3pt}
\end{equation*}%
Then using the UL-type RG-factorization, we obtain%
\begin{equation}
		\setlength{\abovedisplayskip}{3pt}
	\left\{ 
	\begin{array}{l}
		\mathbf{\pi }_{B,0}^{\left( i\right) }=\kappa \mathbf{x}_{0}, \\ 
		\mathbf{\pi }_{B,k}^{\left( i\right) }=\mathbf{\pi }_{B,k-1}^{\left(
			i\right) }R_{k-1,k},\text{ }1\leq k\leq M,%
	\end{array}%
	\right.  \label{eq-iM}
	\setlength{\belowdisplayskip}{3pt}
\end{equation}%
where $\mathbf{x}_{0}$ is the stationary probability vector of the censored
Markov chain $\Psi _{0}$ to level $0$, the scalar $\kappa $ is uniquely
determined by $\sum\nolimits_{k=0}^{M}\mathbf{\pi }_{B,k}^{\left( i\right) }%
\mathbbm{1}=1$. Note that the
censored Markov process $Q^{\left[ \leq 0\right] }$ has the stationary
probability vector $\mathbf{x}_{0}$, and thus we have%
\begin{equation}
		\setlength{\abovedisplayskip}{3pt}
	\left\{ 
	\begin{array}{l}
		\mathbf{x}_{0}Q^{\left[ \leq 0\right] }=0, \\ 
		\mathbf{x}_{0}\mathbbm{1}=1.%
	\end{array}%
	\right.  \label{eq-Q0x}
	\setlength{\belowdisplayskip}{3pt}
\end{equation}

\subsection{Markov process of the maintenance shop}

\label{markov-maintenance}

The repair behavior of the maintenance shop can be analyzed by a
two-dimensional continuous-time Markov process $\left\{ \left( Q_{B}^{\left(
	0\right) }\left( t\right) ,Q_{G}^{\left( 0\right) }\left( t\right) \right)
t\geq 0\right\} $, where $Q_{B}^{\left( 0\right) }\left( t\right) $ is the
phase variable and $Q_{G}^{\left( 0\right) }\left( t\right) $ is the level
variable. When the number of repaired items amounts to $Z$, the $Z$ repaired
items are taken away in batch from the maintenance shop to
bases. We assume that the proportion of repaired items redistributed to base 
$i$ is $\beta _{i}$, where $\beta _{i}=Z_{i}/Z$ for $1\leq i\leq N$ and the
transportation time of the $Z_{i}$ repaired items is exponential with rate $%
\mu _{0,i}$. We write $\mu _{0}=\sum\nolimits_{i=1}^{N}\beta _{i}\mu _{0,i}$%
. Note that there are $r$ repairmen at the maintenance shop, and the repair
time of each failed item is exponential with repair rate $w$. The repair
rate of the maintenance shop is given by%
\begin{equation*}
		\setlength{\abovedisplayskip}{3pt}
	\overline{w}\left( n_{B}^{\left( 0\right) }\right) =\min \left\{
	n_{B}^{\left( 0\right) },r\right\} w,
	\setlength{\belowdisplayskip}{3pt}
\end{equation*}%
where $n_{B}^{\left( 0\right) }$ is the number of failed items in the
maintenance shop. 

Figure 4 depicts the state transition relations of the Markov process $\{(Q_{B}^{\left(  0\right)  }\left(  t\right)  ,Q_{G}^{\left(  0\right)
}\left(  t\right)  ):t\geq0\}$ in the maintenance
shop. Obviously, this Markov process is irreducible and positive recurrent.
The infinitesimal generator $\mathbf{T}$ of the Markov process $\{(Q_{B}^{\left(  0\right)  }\left(  t\right)  ,Q_{G}^{\left(  0\right)
}\left(  t\right)  ):t\geq0\}$ is given by 
\begin{equation}
		\setlength{\abovedisplayskip}{3pt}
	\mathbf{T}=\left( 
	\begin{array}{cccccccc}
		T_{0,0} & T_{0,1} &  &  &  &  &  &  \\ 
		& T_{1,1} & T_{1,2} &  &  &  &  &  \\ 
		&  & \ddots & \ddots &  &  &  &  \\ 
		&  &  & T_{M,M} & T_{M,M+1} &  &  &  \\ 
		&  &  &  & T_{M+1,M+1} & T_{M+1,M+2} &  &  \\ 
		&  &  &  &  & \ddots & \ddots &  \\ 
		&  &  &  &  &  & T_{Z-1,Z-1} & T_{Z-1,Z} \\ 
		T_{Z,0} &  &  &  &  &  &  & T_{Z,Z}%
	\end{array}%
	\right) ,  \label{V-2}
\end{equation}%
whose block elements are given in Appendix D.
\begin{figure}[h!]
	\centering                         \includegraphics[width=14cm]{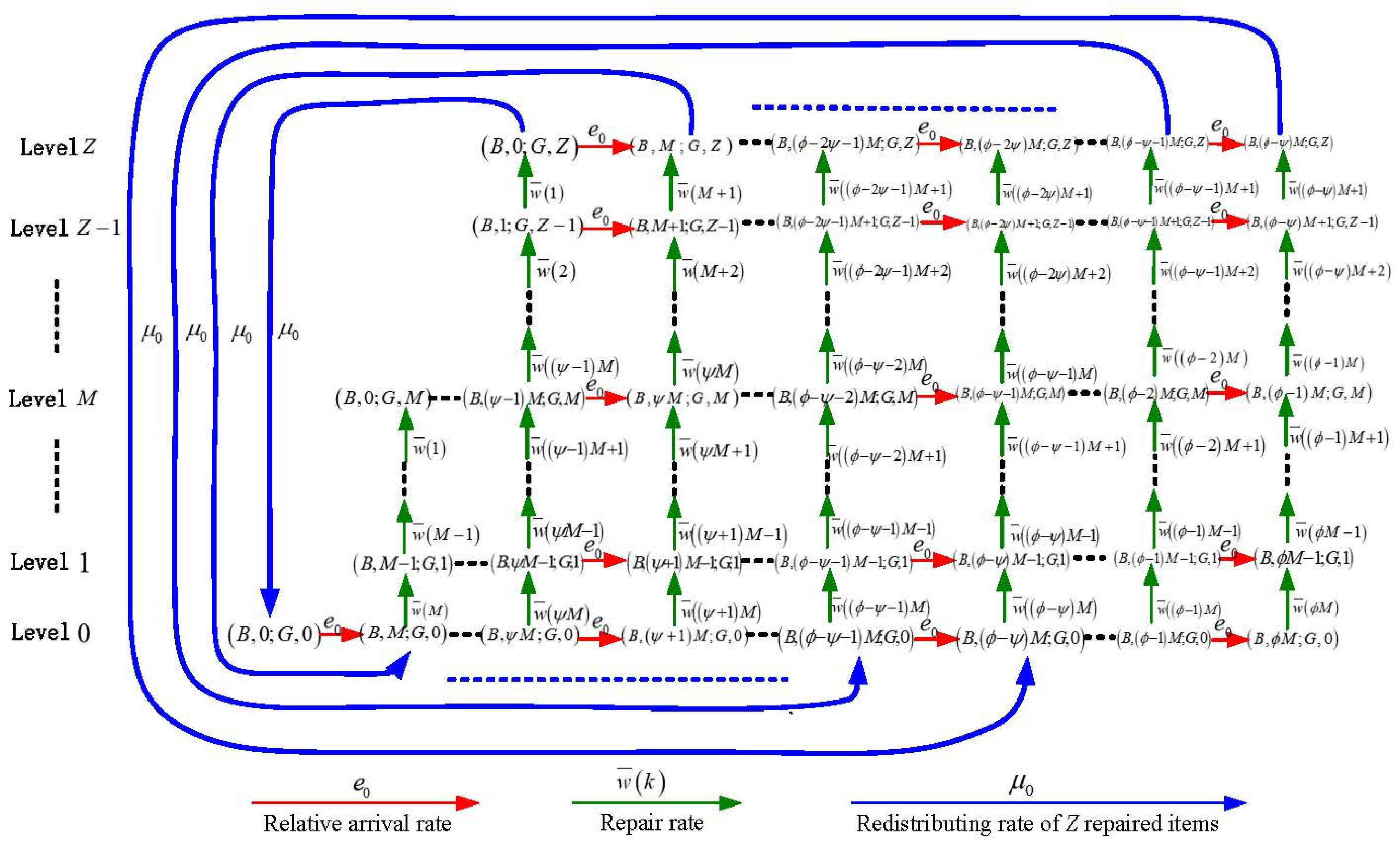} 
	\newline
		\setlength{\abovecaptionskip}{-0.3cm} 
\setlength{\belowcaptionskip}{-0.3cm} 
	\caption{State transition relations of the Markov process in the maintenance
		shop}
	\label{figure:figure-4}
\end{figure}

For the two-dimensional Markov process, we write $%
T_{Z,Z}^{[\leq Z]}=T_{Z,Z}$. For $1\leq k\leq Z-1$, we can iteratively get%
\begin{equation*}
		\setlength{\abovedisplayskip}{3pt}
	T^{[\leq k]}=\left( 
	\begin{array}{ccccc}
		T_{0,0} & T_{0,1} &  &  &  \\ 
		& T_{1,1} & T_{1,2} &  &  \\ 
		&  & \ddots & \ddots &  \\ 
		&  &  & T_{k-1,k-1} & T_{k-1,k} \\ 
		\Xi _{k} &  &  &  & T_{k,k}%
	\end{array}%
	\right) ,
	\setlength{\belowdisplayskip}{3pt}
\end{equation*}%
where $\Xi _{k}=\prod\nolimits_{l=k}^{Z-1}\left[ T_{l,l+1}\left(
-T_{l+1,l+1}\right) ^{-1}\right] T_{Z,0}$.\ Note that $T^{\left[ \leq 0 %
	\right] }=T^{\left[ 0\right] }$, and we obtain%
\begin{equation}
		\setlength{\abovedisplayskip}{3pt}
	T^{\left[ \leq 0\right] }=T_{0,0}+T_{0,1}\left( -T_{1,1}\right) ^{-1}\left[
	\prod\limits_{l=1}^{Z-1}T_{l,l+1}\left( -T_{l+1,l+1}\right) ^{-1}\right]
	T_{Z,0}.  \label{eq-T0}
	\setlength{\belowdisplayskip}{3pt}
\end{equation}

Let%
\begin{equation*}
		\setlength{\abovedisplayskip}{3pt}
	U_{n}=T_{n,n}^{[\leq n]},\text{ }0\leq n\leq Z,
	\setlength{\belowdisplayskip}{3pt}
\end{equation*}%
\begin{equation*}
		\setlength{\abovedisplayskip}{3pt}
	\mathbf{R}_{i,j}=T_{i,j}^{\left[ \leq j\right] }\left( -U_{j}\right) ^{-1},%
	\text{ }0\leq i<j\leq Z,
	\setlength{\belowdisplayskip}{3pt}
\end{equation*}%
and%
\begin{equation*}
		\setlength{\abovedisplayskip}{3pt}
	\mathbf{G}_{i,j}=\left( -U_{i}\right) ^{-1}T_{i,j}^{\left[ \leq i\right] },%
	\text{ }0\leq j<i\leq Z.
	\setlength{\belowdisplayskip}{3pt}
\end{equation*}%
Then the UL-type RG-factorization of the Markov process is
given by
\begin{equation*}
		\setlength{\abovedisplayskip}{3pt}
	\mathbf{T}=\left( I-\mathbf{R}_{U}\right) U_{D}\left( I-\mathbf{G}%
	_{L}\right) ,
	\setlength{\belowdisplayskip}{3pt}
\end{equation*}%
where

\begin{equation*}
	\mathbf{R}_{U}=\left( 
	\begin{array}{ccccccc}
		0 & \mathbf{R}_{0,1} &  &  &  &  &  \\ 
		& 0 & \mathbf{R}_{1,2} &  &  &  &  \\ 
		&  & 0 & \mathbf{R}_{2,3} &  &  &  \\ 
		&  &  & \ddots & \ddots &  &  \\ 
		&  &  &  & 0 & \mathbf{R}_{Z-2,Z-1} &  \\ 
		&  &  &  &  & 0 & \mathbf{R}_{Z-1,Z} \\ 
		&  &  &  &  &  & 0%
	\end{array}%
	\right) ,
\end{equation*}%
\begin{equation*}
		\setlength{\abovedisplayskip}{3pt}
	U_{D}=\text{diag}\left( U_{0},U_{1},\ldots ,U_{Z-1},U_{Z}\right) ,
	\setlength{\belowdisplayskip}{3pt}
\end{equation*}%
and%
\begin{equation*}
		\setlength{\abovedisplayskip}{3pt}
	\mathbf{G}_{L}=\left( 
	\begin{array}{cccccc}
		0 &  &  &  &  &  \\ 
		\mathbf{G}_{1,0} & 0 &  &  &  &  \\ 
		\mathbf{G}_{2,0} &  & 0 &  &  &  \\ 
		\vdots &  &  & \ddots &  &  \\ 
		\mathbf{G}_{Z-1,0} &  &  &  & 0 &  \\ 
		\mathbf{G}_{Z,0} & 0 & \cdots & 0 & 0 & 0%
	\end{array}%
	\right) .
	\setlength{\belowdisplayskip}{3pt}
\end{equation*}

Since the Markov process is irreducible and positive recurrent,
there exists the stationary probability vector. Let $\mathbf{\pi }%
_{G}^{\left( 0\right) }=\left( \pi _{G,0}^{\left( 0\right) },\pi
_{G,1}^{\left( 0\right) },\pi _{G,2}^{\left( 0\right) },\ldots ,\pi
_{G,Z-1}^{\left( 0\right) },\pi _{G,Z}^{\left( 0\right) }\right) $ be the
stationary probability vector of the Markov process, where%
\[
\left\{
\begin{array}
	[l]{l}%
	\mathbf{\pi}_{G,k}^{\left(  0\right)  }=\left(  \pi_{B,0;G,k}^{\left(
		0\right)  },\pi_{B,M;G,k}^{\left(  0\right)  },\ldots,\pi_{B,\left(
		\phi-1\right)  M-k;G,k}^{\left(  0\right)  },\pi_{B,\phi M-k;G,k}^{\left(
		0\right)  }\right)  ,\text{ for }k=lM,\text{ }0\leq l\leq\psi,\\
	\mathbf{\pi}_{G,k}^{\left(  0\right)  }=\left(  \pi_{B,\left(  l+1\right)
		M-k;G,k}^{\left(  0\right)  },\pi_{B,\left(  l+2\right)  M-k;G,k}^{\left(
		0\right)  },\ldots,\pi_{B,\left(  \phi-1\right)  M-k;G,k}^{\left(  0\right)
	},\pi_{B,\phi M-k;G,k}^{\left(  0\right)  }\right)  ,\text{ for }lM+\\
	\text{ \ \ \ \ \ \ \ \ \ }1\leq k<\left(  l+1\right)  M,0\leq l\leq\psi-1.
\end{array}
\right.
\]
Then by using the UL-type RG-factorizations, we obtain%
\begin{equation}
		\setlength{\abovedisplayskip}{3pt}
	\left\{ 
	\begin{array}{l}
		\mathbf{\pi }_{G,0}^{\left( 0\right) }=\sigma \mathbf{y}_{0}, \\ 
		\mathbf{\pi }_{G,k}^{\left( 0\right) }=\mathbf{\pi }_{G,k-1}^{\left(
			0\right) }\mathbf{R}_{k-1,k},\text{ }1\leq k\leq Z,%
	\end{array}%
	\right.  \label{eq-0Z}
	\setlength{\belowdisplayskip}{3pt}
\end{equation}%
where $\mathbf{y}_{0}$ is the stationary probability vector of the censored
Markov process $U_{0}$ to level $0$, and the scalar $\sigma $ is uniquely
determined by $\sum\nolimits_{k=0}^{Z}\mathbf{\pi }_{G,k}^{\left( 0\right) }%
\mathbbm{1}=1$. The censored Markov process $T^{\left[ \leq 0\right] }$ has
the stationary probability vector $\mathbf{y}_{0}$ such that%
\begin{equation}
		\setlength{\abovedisplayskip}{3pt}
	\left\{ 
	\begin{array}{l}
		\mathbf{y}_{0}T^{\left[ \leq 0\right] }=0, \\ 
		\mathbf{y}_{0}\mathbbm{1}=1.%
	\end{array}%
	\right.  \label{eq-T0y}
	\setlength{\belowdisplayskip}{3pt}
\end{equation}

\subsection{The routing matrix}

\label{the-routing}

The routing probabilities of the closed queueing network are complicated since
they depend on the states of each node by considering the items'
failure, removal, repair, redistribution, and reuse processes under two
batch policies. Figure 5 depicts the routing probabilities for the three classes of
nodes. 
\begin{figure}[h!]
	\centering                         \includegraphics[width=8cm]{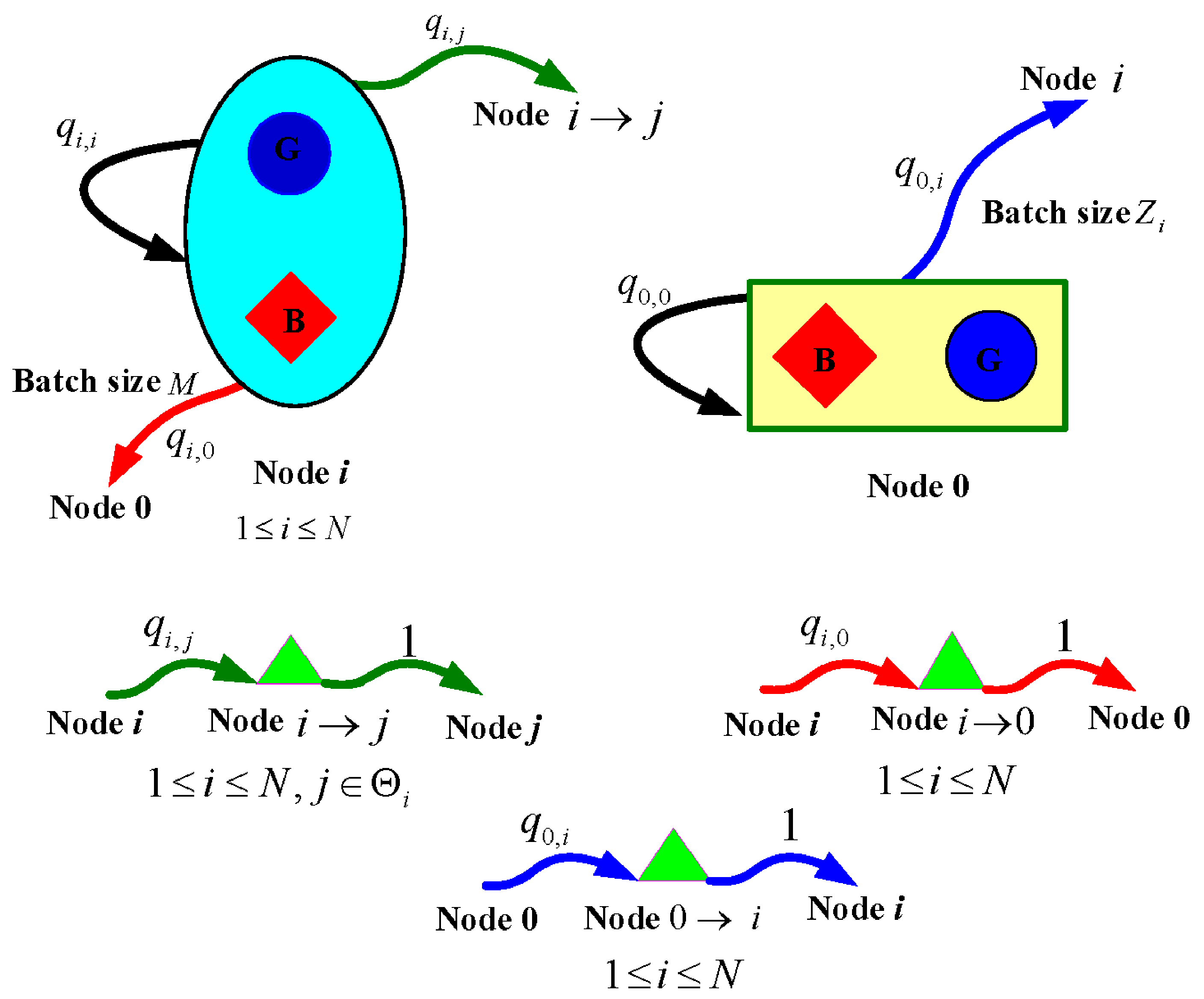} 
	\newline
	\setlength{\abovecaptionskip}{-0.3cm} 
	\setlength{\belowcaptionskip}{-0.3cm} 
	\caption{Routing probabilities of nodes}
	\label{figure:figure-5}
\end{figure}

For the three classes of nodes, we write the routing matrix as $P=\left(
f_{i,j}\right) $, where $f_{i,j}$ is given by

\begin{equation*}%
	f_{i,j}=\left\{ 
	\begin{array}{ll}
		q_{0,i}, & \text{Node }0\text{ to Node }0\rightarrow i\text{, }1\leq i\leq N%
		\text{,} \\ 
		q_{i,0}, & \text{Node }i\text{ to Node }i\rightarrow 0\text{, }1\leq i\leq N%
		\text{,} \\ 
		q_{i,i}, & \text{Node }i\text{ to Node }i\text{, }0\leq i\leq N\text{,} \\ 
		q_{i,j}, & \text{Node }i\text{ to Node }i\rightarrow j\text{, }j\in \Theta
		_{i}\text{, }1\leq i\leq N\text{,} \\ 
		1, & \text{Node }i\rightarrow j\text{ to Node }j\text{, }j\in \Theta _{i}%
		\text{, }1\leq i\leq N\text{,} \\ 
		1, & \text{Node }i\rightarrow 0\text{ to Node }0\text{, }1\leq i\leq N\text{,%
		} \\ 
		1, & \text{Node }0\rightarrow i\text{ to Node }i\text{, }1\leq i\leq N\text{,%
		} \\ 
		0, & \text{otherwise,}%
	\end{array}%
	\right.
	\setlength{\belowdisplayskip}{3pt}
\end{equation*}%
where%
\begin{equation*}
		\setlength{\abovedisplayskip}{3pt}
	q_{0,0}=1-\left( \sum\limits_{l=0}^{\phi -\psi }\pi _{B,lM;G,Z}^{\left(
		0\right) }\right) ,
	\setlength{\belowdisplayskip}{3pt}
\end{equation*}%
for $1\leq i\leq N$%
\begin{equation*}
		\setlength{\abovedisplayskip}{3pt}
	q_{0,i}=\left( \sum\limits_{l=0}^{\phi -\psi }\pi _{B,lM;G,Z}^{\left(
		0\right) }\right) \beta _{i},\text{ }q_{i,0}=\sum\limits_{k=0}^{K-M}\pi
	_{G,k;B,M}^{\left( i\right) },\text{ }q_{i,i}=\sum\limits_{k=0}^{M-1}\pi
	_{G,0;B,k}^{\left( i\right) },
	\setlength{\belowdisplayskip}{3pt}
\end{equation*}%
and for $1\leq i\leq N,$ $j\in \Theta _{i}$,%
\begin{equation*}
	q_{i,j}=\left( 1-q_{i,0}-q_{i,i}\right) p_{i,j}.
	\setlength{\belowdisplayskip}{3pt}
\end{equation*}%
Note that the routing probabilities are expressed by the stationary
probabilities of the two Markov processes $\left\{
\left( Q_{G}^{\left( i\right) }\left( t\right) ,Q_{B}^{\left( i\right)
}\left( t\right) \right) :t\geq 0\right\} $ and $\{(Q_{B}^{\left(  0\right)  }\left(  t\right)  ,Q_{G}^{\left(  0\right)
}\left(  t\right)  ):t\geq0\}$ ,
both of which are determined by the relative arrival rate vector $\mathbf{e}$%
. Thus the routing matrix depends on $\mathbf{e}$ and can be written as $%
P\left( \mathbf{e}\right) $.

Then the relative arrival rate vector $\mathbf{e}$
satisfies a nonlinear routing matrix equation 
\begin{equation}
		\setlength{\abovedisplayskip}{3pt}
	\mathbf{e}=\mathbf{e}P\left( \mathbf{e}\right) .  \label{ep}
	\setlength{\belowdisplayskip}{3pt}
\end{equation}%
To compute the relative arrival rates, we develop the Algorithm 1 to give an
iterative approximate solution to Equation (\ref{ep}). 

\begin{algorithm}[h]
	\caption{An Iterative Algorithm for Computing the Relative Arrival Rates}\label{alg:1}%
	
	\begin{algorithmic}
		\State \textbf{Step 0: Initialization}
		
		Give an initial vector $_{0}\mathbf{e}$.
		
		\State \textbf{Step 1: The first iterative computation}
		
		(a) Using $_{0}\mathbf{e}$, compute stationary probability vector $_{0}%
		\mathbf{\pi }_{B}^{\left( i\right) }$ for $1\leq i\leq N$ and $_{0}\mathbf{%
			\pi }_{G}^{\left( 0\right) }$ by UL-type RG-factorization.
		
		(b) Obtain the routing matrix $P\left( _{0}\mathbf{e}%
		\right) $ based on $_{0}\mathbf{\pi }_{B}^{\left( i\right) }$ and $_{0}\mathbf{\pi
		}_{G}^{\left( 0\right) }$.
		
		(c) Obtain $_{1}\mathbf{e}$ by the equation $_{1}\mathbf{e}= {_{0}\mathbf{e}}%
		P\left( _{0}\mathbf{e}\right) $.
		
		\State \textbf{Step 2: The second iterative computation}
		
		(a) Using $_{1}\mathbf{e}$, compute stationary probability vector $_{1}\pi
		_{B}^{\left( i\right) }$ for $1\leq i\leq N$ and $_{1}\mathbf{\pi }%
		_{G}^{\left( 0\right) }$ by UL-type RG-factorization.
		
		(b) Obtain routing matrix $P\left( _{1}\mathbf{e}\right)
		$ based on $_{1}\mathbf{\pi }_{B}^{\left( i\right) }$ and $_{1}\mathbf{\pi
		}_{G}^{\left( 0\right) }$.
		
		(c) Obtain $_{2}\mathbf{e}$ by the equation $_{2}\mathbf{e}= {_{1}\mathbf{e}}%
		P\left( _{1}\mathbf{e}\right) $.
		
		\State\textbf{Step 3: The $(k+1)$st iterative computation for $k\geq 2$}
		
		(a) Using $_{k}\mathbf{e}$, compute stationary probability vector $_{k}%
		\mathbf{\pi }_{B}^{\left( i\right) }$ for $1\leq i\leq N$ and $_{k}\mathbf{%
			\pi }_{G}^{\left( 0\right) }$ by UL-type RG-factorization.
		
		(b) Obtain routing matrix $P\left( _{k}\mathbf{e}\right) $ based on $_{k}%
		\mathbf{\pi }_{B}^{\left( i\right) }$ and $_{k}\mathbf{\pi }_{G}^{\left(
			0\right) }$.
		
		(c) Obtain $_{k+1}\mathbf{e}$ by the equation $_{k+1}\mathbf{e}= {_{k}%
			\mathbf{e}}P\left( _{k}\mathbf{e}\right) $.
		
		\State\textbf{Step 4: Convergence check}
		
		If there exists a relative arrival rate vector $_{n}\mathbf{e}$ such that%
		\begin{equation*}
				\setlength{\abovedisplayskip}{3pt}
			\sqrt{\left( {_{n+1}\mathbf{e}_{0}-}\text{ }_{n}\mathbf{e}_{0}\right)
				^{2}+\left( {{_{n+1}\mathbf{e}_{1}-}}\text{ }_{n}\mathbf{e}_{1}\right)
				^{2}+\cdots +\left( {_{n+1}\mathbf{e}_{N}-}\text{ }_{n}\mathbf{e}_{N}\right)
				^{2}}<\varepsilon ,
			\setlength{\belowdisplayskip}{3pt}
		\end{equation*}%
		(called a stop condition), for a given precision $\varepsilon =1e-10$, then
		the computation stops. In this case, $_{n+1}\mathbf{e}\approx $ $_{n}\mathbf{%
			e}$. Otherwise, return to \textbf{Step 3} until this stop condition is
		satisfied.
		
		\State \textbf{Step 5: Output}
		
		Obtain the relative arrival rate vector $_{n}\mathbf{e}$.
	\end{algorithmic}
\end{algorithm}

\section{A Product-Form Solution and Performance Analysis}

\label{a-product}

In this section, we obtain the product-form solution for the stationary
joint probabilities of queue lengths in the closed queueing network. Based
on this, we can define and compute some useful performance measures of the
maintenance network.

\subsection{The product-form solution}

\label{the-solution}

The numbers of usable items and failed items in any base and in the maintenance shop are described as block-structured
Markov processes, and thus the stationary probabilities of Markov processes
of the two classes of nodes can be computed by the UL-type RG-factorization.

For the closed queueing network, Theorem 2 provides the
product-form solution of the stationary joint probabilities $\pi \left( 
\mathbf{n}\right) $ of queue lengths for $\mathbf{n}\in \Omega $.

\begin{The}
	\label{The:PS} For the closed queueing network of the maintenance network, the stationary joint probability $\pi \left( 
	\mathbf{n}%
	\right) $ is given by 
	\begin{equation*}
			\setlength{\abovedisplayskip}{3pt}
		\pi \left( \mathbf{n}\right) =\frac{1}{\mathbf{C}}\prod\limits_{i=0}^{N}H
		\left( n_{G}^{\left( i\right) },n_{B}^{\left( i\right) }\right) \prod\limits 
		_{\substack{ i=1  \\ j\in \Theta _{i}}}^{N}H\left( m_{i,j}\right)
		\prod\limits_{i=1}^{N}H\left( m_{i,0}\right) H\left( m_{0,i}\right) ,
		\setlength{\belowdisplayskip}{3pt}
	\end{equation*}
where%
\[
H\left(  n_{G}^{\left(  i\right)  },n_{B}^{\left(  i\right)  }\right)
=\left\{
\begin{array}
	[c]{ll}%
	\pi_{G,n_{G}^{\left(  i\right)  };B,n_{B}^{\left(  i\right)  }}^{\left(
		i\right)  }, & 1\leq i\leq N,0\leq n_{G}^{\left(  i\right)  }\leq K,0\leq
	n_{B}^{\left(  i\right)  }\leq M,\\
	\pi_{B,n_{B}^{\left(  0\right)  };G,n_{G}^{\left(  0\right)  }}^{\left(
		0\right)  }, & \text{ }i=0,0\leq n_{B}^{\left(  0\right)  }\leq Z,0\leq
	n_{G}^{\left(  0\right)  }\leq\phi M,
\end{array}
\right.
\]%

\[
H\left(  m_{i,j}\right)  =\left(  \frac{e_{i,j}}{\mu_{i,j}}\right)  ^{m_{i,j}%
}\frac{1}{m_{i,j}!},0\leq m_{i,j}\leq K,
\]%
\[
H\left(  m_{i,0}\right)  =\left(  \frac{e_{i,0}}{\mu_{i,0}}\right)  ^{m_{i,0}%
}\frac{1}{m_{i,0}!},\text{ }m_{i,0}=kM\text{ for }0\leq k\leq\phi,
\]%
\[
H\left(  m_{0,i}\right)  =\left(  \frac{e_{0,i}}{\mu_{0,i}}\right)  ^{m_{0,i}%
}\frac{1}{\left(  m_{0,i}\right)  !},\text{ }m_{0,i}=lZ_{i}\text{ for }0\leq
l\leq\frac{\phi}{\psi},
\]
and $\mathbf{C}$ is a normalization constant, given by%
	\begin{equation*}
			\setlength{\abovedisplayskip}{3pt}
		\mathbf{C}=\sum\limits_{\mathbf{n}\in \Omega }\prod\limits_{i=0}^{N}H\left(
		n_{G}^{\left( i\right) },n_{B}^{\left( i\right) }\right) \prod\limits 
		_{\substack{ i=1  \\ j\in \Theta _{i}}}^{N}H\left( m_{i,j}\right)
		\prod\limits_{i=1}^{N}H\left( m_{i,0}\right) H\left( m_{0,i}\right) .
		\setlength{\belowdisplayskip}{3pt}
	\end{equation*}
\end{The}

The proof is given in Appendix E.

The following theorem establishes relations between the marginal
stationary probabilities and the joint stationary probabilities.
\begin{The}
	\label{The:DN} For the closed queueing network, the marginal probabilities
	of the system can be determined by the joint stationary probabilities as
	follows:
	
	(1) For Node $i,1\leq i\leq N,0\leq l\leq M,0\leq k\leq K-l,$ 
	\begin{equation*}
			\setlength{\abovedisplayskip}{3pt}
		\pi \left( \mathbf{n:}n_{G}^{\left( i\right) }=k,n_{B}^{\left( i\right)
		}=l\right) =\sum\limits_{\substack{ \mathbf{n}\in \Omega  \\ n_{G}^{\left(
					i\right) }=k,n_{B}^{\left( i\right) }=l}}\pi \left( \mathbf{n}\right) =\pi
		_{G,k;B,l}^{\left( i\right) }\frac{\widetilde{\mathbf{C}}\left(
			n_{G}^{\left( i\right) }=k,n_{B}^{\left( i\right) }=l\right) }{\mathbf{C}},
		\setlength{\belowdisplayskip}{3pt}
	\end{equation*}
	where%
	\begin{equation*}
			\setlength{\abovedisplayskip}{3pt}
		\widetilde{\mathbf{C}}\left( n_{G}^{\left( i\right) }=k,n_{B}^{\left(
			i\right) }=l\right) =\sum\limits_{\substack{ \mathbf{n}\in \Omega  \\ %
				n_{G}^{\left( i\right) }=k,n_{B}^{\left( i\right) }=l}}\prod\limits 
		_{\substack{ j=0  \\ j\neq i}}^{N}H\left( n_{G}^{\left( j\right)
		},n_{B}^{\left( j\right) }\right) \prod\limits_{\substack{ j=1  \\ h\in
				\Theta _{j}}}^{N}H\left( m_{j,h}\right) \prod\limits_{j=1}^{N}H\left(
		m_{j,0}\right) H\left( m_{0,j}\right) .
		\setlength{\belowdisplayskip}{3pt}
	\end{equation*}
	
	(2) For Node $i,i=0,0\leq k\leq Z,k+l=hM,0\leq h\leq \phi ,$ 
	\begin{equation*}
			\setlength{\abovedisplayskip}{3pt}
		\pi \left( \mathbf{n:}n_{G}^{\left( 0\right) }=k,n_{B}^{\left( 0\right)
		}=l\right) =\sum\limits_{\substack{ \mathbf{n}\in \Omega  \\ n_{G}^{\left(
					0\right) }=k,n_{B}^{\left( 0\right) }=l}}\pi \left( \mathbf{n}\right) =\pi
		_{B,l;G,k}^{\left( 0\right) }\frac{\widetilde{\mathbf{C}}\left(
			n_{G}^{\left( 0\right) }=k,n_{B}^{\left( 0\right) }=l\right) }{\mathbf{C}},
		\setlength{\belowdisplayskip}{3pt}
	\end{equation*}
	where%
	\begin{equation*}
			\setlength{\abovedisplayskip}{3pt}
		\widetilde{\mathbf{C}}\left( n_{G}^{\left( 0\right) }=k,n_{B}^{\left(
			0\right) }=l\right) =\sum\limits_{\substack{ \mathbf{n}\in \Omega  \\ %
				n_{G}^{\left( 0\right) }=k,n_{B}^{\left( 0\right) }=l}}\prod
		\limits_{j=1}^{N}H\left( n_{G}^{\left( j\right) },n_{B}^{\left( j\right)
		}\right) \prod\limits_{\substack{ j=1  \\ h\in \Theta _{j}}}^{N}H\left(
		m_{j,h}\right) \prod\limits_{j=1}^{N}H\left( m_{j,0}\right) H\left(
		m_{0,j}\right) .
		\setlength{\belowdisplayskip}{3pt}
	\end{equation*}
	
	(3) For Node $i\rightarrow j,1\leq i\leq N,j\in \Theta _{i},0\leq h\leq K,$ 
	\begin{equation*}
			\setlength{\abovedisplayskip}{3pt}
		\pi \left( \mathbf{n:}m_{i,j}=h\right) =\sum\limits_{\substack{ \mathbf{n}
				\in \Omega  \\ m_{i,j}=h}}\pi \left( \mathbf{n}\right) =\frac{1}{h!}\left( 
		\frac{e_{i,j}}{\mu _{i,j}}\right) ^{h}\frac{\widetilde{\mathbf{C}}\left(
			m_{i,j}=h\right) }{\mathbf{C}},
		\setlength{\belowdisplayskip}{3pt}
	\end{equation*}
	where%
	\begin{equation*}
			\setlength{\abovedisplayskip}{3pt}
		\widetilde{\mathbf{C}}\left( m_{i,j}=h\right) =\sum\limits_{\substack{ 
				\mathbf{n}\in \Omega  \\ m_{i,j}=h}}\prod\limits_{k=0}^{N}H\left(
		n_{G}^{\left( k\right) },n_{B}^{\left( k\right) }\right) \prod\limits 
		_{\substack{ k=1  \\ l\in \Theta _{k}  \\ \left( k,l\right) \neq \left(
				i,j\right) }}^{N}H\left( m_{k,l}\right) \prod\limits_{k=1}^{N}H\left(
		m_{k,0}\right) H\left( m_{0,k}\right) .
		\setlength{\belowdisplayskip}{3pt}
	\end{equation*}
	
	(4) For Node $i\rightarrow 0$, $1\leq i\leq N,0\leq h\leq \phi ,$ 
	\begin{equation*}
			\setlength{\abovedisplayskip}{3pt}
		\pi \left( \mathbf{n:}m_{i,0}=hM\right) =\sum\limits_{\substack{ \mathbf{n}
				\in \Omega  \\ m_{i,0}=hM}}\pi \left( \mathbf{n}\right) =\frac{1}{\left(
			hM\right) !}\left( \frac{e_{i,0}}{\mu _{i,0}}\right) ^{hM}\frac{\widetilde{%
				\mathbf{C}}\left( m_{i,0}=hM\right) }{\mathbf{C}},
			\setlength{\belowdisplayskip}{3pt}
	\end{equation*}
	where%
	\begin{equation*}
			\setlength{\abovedisplayskip}{3pt}
		\widetilde{\mathbf{C}}\left( m_{i,0}=hM\right) =\sum\limits_{\substack{ 
				\mathbf{n}\in \Omega  \\ m_{i,0}=hM}}\prod\limits_{k=0}^{N}H\left(
		n_{G}^{\left( k\right) },n_{B}^{\left( k\right) }\right) \prod\limits 
		_{\substack{ k=1  \\ l\in \Theta _{k}}}^{N}H\left( m_{k,l}\right)
		\prod\limits _{\substack{ k=1  \\ k\neq i}}^{N}H\left( m_{k,0}\right)
		\prod\limits_{k=1}^{N}H\left( m_{0,k}\right) .
		\setlength{\belowdisplayskip}{3pt}
	\end{equation*}
	
	(5) For Node $0\rightarrow i,1\leq i\leq N,0\leq l\leq \phi /\psi ,$ 
	\begin{equation*}
			\setlength{\abovedisplayskip}{3pt}
		\pi \left( \mathbf{n:}m_{0,i}=lZ_{i}\right) =\sum\limits_{\substack{ \mathbf{%
					\ n}\in \Omega  \\ m_{0,i}=lZ_{i}}}\pi \left( \mathbf{n}\right) =\frac{1}{%
			\left( lZ_{i}\right) !}\left( \frac{e_{0,i}}{\mu _{0,i}}\right) ^{lZ_{i}} 
		\frac{\widetilde{\mathbf{C}}\left( m_{0,i}=lZ_{i}\right) }{\mathbf{C}},
		\setlength{\belowdisplayskip}{3pt}
	\end{equation*}
	where%
	\begin{equation*}
			\setlength{\abovedisplayskip}{3pt}
		\widetilde{\mathbf{C}}\left( m_{0,i}=lZ_{i}\right) =\sum\limits_{\substack{ 
				\mathbf{n}\in \Omega  \\ m_{0,i}=lZ_{i}}}\prod\limits_{k=0}^{N}H\left(
		n_{G}^{\left( k\right) },n_{B}^{\left( k\right) }\right) \prod\limits 
		_{\substack{ k=1  \\ l\in \Theta _{k}}}^{N}H\left( m_{k,l}\right)
		\prod\limits_{k=1}^{N}H\left( m_{k,0}\right) \prod\limits_{\substack{ k=1 
				\\ k\neq i}}^{N}H\left( m_{0,k}\right) .
			\setlength{\belowdisplayskip}{3pt}
	\end{equation*}
\end{The}

The proof is immediate by Section 7 in \cite%
{bolch2006queueing}, and hence it is omitted here. 

\subsection{Performance Analysis}

\label{performance-analysis}

In this subsection, we provide performance measures for the
maintenance network by using the stationary joint
probability vector $\pi \left( \mathbf{n}\right) $ for $\mathbf{n}\in \Omega 
$.

\textbf{(a) The stationary proportion of failed items in the maintenance
	network}

In the maintenance network, we are interested in the stationary
proportion of failed items that are distributed in
the $N$ bases, the maintenance shop, and the path nodes from bases to the maintenance shop. The average number of 
failed items $E%
\left[ \Im \right] $ can be computed by the marginal probabilities
as follows:%
\begin{equation*}
		\setlength{\abovedisplayskip}{3pt}
	E\left[ \Im \right] =\sum\limits_{i=1}^{N}\sum\limits_{l=0}^{M}l\mathbf{\pi }%
	\left( \mathbf{n:}n_{B}^{\left( i\right) }=l\right) +\sum\limits_{l=0}^{\phi
		M}l\mathbf{\pi }\left( \mathbf{n:}n_{B}^{\left( 0\right) }=l\right)
	+\sum\limits_{i=1}^{N}\sum\limits_{l=0}^{\phi }lM\mathbf{\pi }\left( \mathbf{%
		\ n:}m_{i,0}=lM\right) ,
	\setlength{\belowdisplayskip}{3pt}
\end{equation*}%
where the three terms are the average numbers of failed items in
the $N$ bases, the maintenance shop, and Node $i\rightarrow 0$ for $1\leq
i\leq N$, respectively. Based on Theorem 3, we obtain%
\begin{equation*}
		\setlength{\abovedisplayskip}{3pt}
	\mathbf{\pi }\left( \mathbf{n:}n_{B}^{\left( i\right) }=l\right) =\frac{1}{%
		\mathbf{C}}\sum\limits_{k=0}^{K-l}\left[ \pi _{G,k;B,l}^{\left( i\right) }%
	\widetilde{\mathbf{C}}\left( n_{G}^{\left( i\right) }=k,n_{B}^{\left(
		i\right) }=l\right) \right] ,
	\setlength{\belowdisplayskip}{3pt}
\end{equation*}%
\begin{equation*}
		\setlength{\abovedisplayskip}{3pt}
	\mathbf{\pi }\left( \mathbf{n:}n_{B}^{\left( 0\right) }=l\right) =\frac{1}{%
		\mathbf{C}}\sum\limits_{h=0}^{\phi }\sum\limits_{k=0}^{Z}\sum\limits_{k+l=hM}%
	\left[ \pi _{B,l;G,k}^{\left( 0\right) }\widetilde{\mathbf{C}}\left(
	n_{G}^{\left( 0\right) }=k,n_{B}^{\left( 0\right) }=l\right) \right] ,
	\setlength{\belowdisplayskip}{3pt}
\end{equation*}%
and%
\begin{equation*}
		\setlength{\abovedisplayskip}{3pt}
	\mathbf{\pi }\left( \mathbf{n:}m_{i,0}=lM\right) =\frac{1}{\left( lM\right) !%
	}\left( \frac{e_{i,0}}{\mu _{i,0}}\right) ^{lM}\frac{\widetilde{\mathbf{C}}%
		\left( m_{i,0}=lM\right) }{\mathbf{C}}.
	\setlength{\belowdisplayskip}{3pt}
\end{equation*}%
Thus the proportion of failed items in the maintenance network
with respect to the total number $K$ of items in the maintenance network is
given by%
\begin{equation*}
		\setlength{\abovedisplayskip}{3pt}
	\eta =\frac{E\left[ \Im \right] }{K}.
	\setlength{\belowdisplayskip}{3pt}
\end{equation*}

\textbf{(b)} \textbf{The stationary proportion of usable items in either
	various bases or the path nodes}

The path nodes with usable items have two different cases. We denote the
average number $E\left[ \varpi \right] $ of usable items in
either the bases or the path nodes by the
marginal probabilities as follows:
\begin{align*}
	\setlength{\abovedisplayskip}{3pt} 
	E\left[  \varpi\right]     =&\sum
	\limits_{i=1}^{N}\sum\limits_{k=0}^{K}k\mathbf{\ \pi}\left(  \mathbf{n:}%
	n_{G}^{\left(  i\right)  }=k\right)  +\sum\limits_{i=1}^{N}\sum\limits_{j\in
		\Theta_{i}}\sum\limits_{k=0}^{K}k\mathbf{\pi}\left(  \mathbf{n:}%
	m_{i,j}=k\right)  \\
	& +\sum\limits_{i=1}^{N}\sum\limits_{k=0}^{\phi/\psi}k\beta_{i}Z\mathbf{\pi
	}\left(  \mathbf{n:}m_{0,i}=k\beta_{i}Z\right)  ,
\setlength{\belowdisplayskip}{3pt}
\end{align*}
where the three terms are the average number of usable items in
the $N$\ bases, the Node $i\rightarrow j$, and Node $0\rightarrow i$ for $%
1\leq i\leq N$ and $j\in \Theta _{i}$, respectively. Based on Theorem 3, we
obtain%
\begin{equation*}
		\setlength{\abovedisplayskip}{3pt}
	\mathbf{\pi }\left( \mathbf{n:}n_{G}^{\left( i\right) }=k\right) =\frac{1}{%
		\mathbf{C}}\sum\limits_{l=0}^{\min \left( M,K-k\right) }\left[ \pi
	_{G,k;B,l}^{\left( i\right) }\widetilde{\mathbf{C}}\left( n_{G}^{\left(
		i\right) }=k,n_{B}^{\left( i\right) }=l\right) \right] ,
	\setlength{\belowdisplayskip}{3pt}
\end{equation*}%
\begin{equation*}
		\setlength{\abovedisplayskip}{3pt}
	\mathbf{\pi }\left( \mathbf{n:}m_{i,j}=k\right) =\frac{1}{k!}\left( \frac{%
		e_{i,j}}{\mu _{i,j}}\right) ^{k}\frac{\widetilde{\mathbf{C}}\left(
		m_{i,j}=k\right) }{\mathbf{C}},
\end{equation*}%
and%
\begin{equation*}
		\setlength{\abovedisplayskip}{3pt}
	\mathbf{\pi }\left( \mathbf{n:}m_{0,i}=k\beta _{i}Z\right) =\frac{1}{\left(
		k\beta _{i}Z\right) !}\left( \frac{e_{0,i}}{\mu _{0,i}}\right) ^{k\beta
		_{i}Z}\frac{\widetilde{\mathbf{C}}\left( m_{0,i}=k\beta _{i}Z\right) }{%
		\mathbf{C}}.
	\setlength{\belowdisplayskip}{3pt}
\end{equation*}%
Thus the stationary proportion of usable items in either bases or the path nodes with respect to the total number $K$ is given by%
\begin{equation*}
		\setlength{\abovedisplayskip}{3pt}
	\xi =\frac{E\left[ \varpi \right] }{K}.
	\setlength{\belowdisplayskip}{3pt}
\end{equation*}%
Note that the stationary proportion $\xi $ measures the availability of
items that can directly serve the arriving users.

\textbf{(c) The busy probability of the maintenance shop}

Let $F_{0}$ be the stationary probability that there is no failed item
in the maintenance shop. Then, we have%
\begin{equation*}
		\setlength{\abovedisplayskip}{3pt}
	F_{0}=\mathbf{\pi }\left( \mathbf{n:}n_{B}^{\left( 0\right) }=0\right) =%
	\frac{1}{\mathbf{C}}\sum\limits_{k=0}^{\psi }\left[ \pi _{B,0;G,kM}^{\left(
		0\right) }\widetilde{\mathbf{C}}\left( n_{G}^{\left( 0\right)
	}=kM,n_{B}^{\left( 0\right) }=0\right) \right] ,
\setlength{\belowdisplayskip}{3pt}
\end{equation*}%
where%
\begin{align*}
	\setlength{\abovedisplayskip}{3pt}
	&  \widetilde{\mathbf{C}}\left(
	n_{G}^{\left(  0\right)  }=kM,n_{B}^{\left(  0\right)  }=0\right)  \\
	& =\sum\limits_{\substack{\mathbf{n}\in\Omega\\n_{G}^{\left(  0\right)
			}=kM,n_{B}^{\left(  0\right)  }=0}}\prod\limits_{j=1}^{N}H\left(
	n_{G}^{\left(  j\right)  },n_{B}^{\left(  j\right)  }\right)  \prod
	\limits_{\substack{j=1\\h\in\Theta_{j}}}^{N}H\left(  m_{j,h}\right)
	\prod\limits_{j=1}^{N}H\left(  m_{j,0}\right)  H\left(  m_{0,j}\right)	.
	\setlength{\belowdisplayskip}{3pt}
\end{align*}
Let 
\begin{equation*}
		\setlength{\abovedisplayskip}{3pt}
	F_{A}=1-F_{0}.
	\setlength{\belowdisplayskip}{3pt}
\end{equation*}%
Then the probability $F_{A}$ is the busy probability of the
maintenance shop, which can be used to measure the maintenance process and the sizes of the two batch policies.

\textbf{(d) Two useful stationary proportions}

Let $\gamma _{1}$ be the proportion of the repaired items with respect to
all the items in the maintenance shop. Then%
\begin{equation*}
		\setlength{\abovedisplayskip}{3pt}
	\gamma _{1}=\frac{\sum\limits_{k=0}^{Z}k\mathbf{\pi }\left( \mathbf{n:}
		n_{G}^{\left( 0\right) }=k\right) }{\sum\limits_{h=0}^{\phi
		}\sum\limits_{k+l=hM}(k+l)\mathbf{\pi }\left( \mathbf{n:}n_{G}^{\left(
			0\right) }=k,n_{B}^{\left( 0\right) }=l\right) },
		\setlength{\belowdisplayskip}{3pt}
\end{equation*}%
where%
\begin{equation*}
		\setlength{\abovedisplayskip}{3pt}
	\mathbf{\pi }\left( \mathbf{n:}n_{G}^{\left( 0\right) }=k,n_{B}^{\left(
		0\right) }=l\right) =\frac{1}{\mathbf{C}}\pi _{B,l;G,k}^{\left( 0\right) }%
	\widetilde{\mathbf{C}}\left( n_{G}^{\left( 0\right) }=k,n_{B}^{\left(
		0\right) }=l\right) ,
	\setlength{\belowdisplayskip}{3pt}
\end{equation*}%
and%
\begin{equation*}
		\setlength{\abovedisplayskip}{3pt}
	\mathbf{\pi }\left( \mathbf{n:}n_{G}^{\left( 0\right) }=k\right) =\frac{1}{%
		\mathbf{C}}\sum\limits_{h=0}^{\phi }\sum\limits_{l=hM-k}\left[ \pi
	_{B,l;G,k}^{\left( 0\right) }\widetilde{\mathbf{C}}\left( n_{G}^{\left(
		0\right) }=k,n_{B}^{\left( 0\right) }=l\right) \right] .
	\setlength{\belowdisplayskip}{3pt}
\end{equation*}

Let $\gamma _{2}$ denote the proportion of the failed items in the
maintenance shop with respect to all the failed items in the maintenance
network 
\begin{equation*}
		\setlength{\abovedisplayskip}{3pt}
	\gamma _{2}=\frac{\sum\limits_{l=0}^{\phi M}l\mathbf{\pi }\left( \mathbf{n:}
		n_{B}^{\left( 0\right) }=l\right) }{E\left[ \Im \right] },
	\setlength{\belowdisplayskip}{3pt}
\end{equation*}%
where%
\begin{equation*}
		\setlength{\abovedisplayskip}{3pt}
	\mathbf{\pi }\left( \mathbf{n:}n_{B}^{\left( 0\right) }=l\right) =\frac{1}{%
		\mathbf{C}}\sum\limits_{h=0}^{\phi }\sum\limits_{k=hM-l}\left[ \pi
	_{B,l;G,k}^{\left( 0\right) }\widetilde{\mathbf{C}}\left( n_{G}^{\left(
		0\right) }=k,n_{B}^{\left( 0\right) }=l\right) \right] .
		\setlength{\belowdisplayskip}{3pt}
\end{equation*}

The two proportions $\gamma _{1}$ and $\gamma _{2}$ can be used to
measure the repair ability of the maintenance shop and the removal ability
of failed items in the maintenance network, respectively.

The computational method developed in this paper and the two batch policies
can be applied to improve the quality of service, system design, and
operations management of the maintenance network.

\section{Concluding Remarks}

\label{concluding}

This paper studies a maintenance network and develops a
new computational method by applying the RG-factorizations of
block-structured Markov processes in the closed queueing network. First, we
describe the maintenance network as a closed
queueing network that characterizes the items' failure, removal,
redistribution, and reuse processes under two batch policies. Then we show
that the relative arrival rates satisfy a nonlinear routing matrix equation.
The Markov systems in any base and the maintenance shop are treated by
the block-structured Markov processes whose stationary probability vectors
can be computed by the RG-factorizations. Finally, we provide a more general
product-form solution for the closed queueing network by extending and
generalizing from simple queueing systems of the nodes to general
block-structured Markov processes. Based on this, we provide performance measures of the maintenance network with failed items.

We hope the methodology and results given in this paper open a new research direction on the closed queueing network 
and
 shed
light on the analysis of more general maintenance networks.
Along the research line, there are still a number of interesting directions
for future research. First, it is significant to
	develop periodic policies both for removing the failed items and
	redistributing the repaired items. The second issue is to develop new algorithms to solve the nonlinear matrix
	equation system satisfied by the relative arrival rates. Third, the optimization problem of the batch policies for the maintenance 
	network are worth further exploration.

\section*{Acknowledgements}
Quan-Lin Li is supported by the National Natural Science Foundation of China
under grant Nos. 71671158 and 71932002, and by the Natural Science
Foundation of Hebei province under grant No. G2017203277. Xiaole Wu is supported by the National Natural Science Foundation of 
China under grant Nos. 71622001, 91746302, and 72025102.

	\section*{Appendices}

\section*{Appendix A. The elements of matrix $\mathbf{Q}$}

We provide expression for the block element $Q_{m,n}$ of matrix $%
\mathbf{Q}$ in Equation (1), where $m$ and $n$ are two level
variables with $0\leq m,n\leq M$. We consider base $i$. For $0\leq n\leq M-1$ 
\begin{equation*}
	\setlength{\abovedisplayskip}{3pt}
	Q_{n,n}=\left( 
	\begin{array}{ccccc}
		-e_{i} & e_{i} &  &  &  \\ 
		\lambda _{i} & -\left( \lambda _{i}+e_{i}+\alpha \right) & e_{i} &  &  \\ 
		& \ddots & \ddots &  &  \\ 
		&  & \lambda _{i} & -\left( \lambda _{i}+e_{i}+\left( K-n-1\right) \alpha
		\right) & e_{i} \\ 
		&  &  & \lambda _{i} & -\left( \lambda _{i}+\left( K-n\right) \alpha \right)%
	\end{array}
	\right) ,
	\setlength{\belowdisplayskip}{3pt}
\end{equation*}%
whose size is $\left( K-n+1\right) \times \left( K-n+1\right) $, and 
\begin{equation*}
	\setlength{\abovedisplayskip}{3pt}
	Q_{n,n+1}=\left( 
	\begin{array}{ccccc}
		0 &  &  &  &  \\ 
		\alpha &  &  &  &  \\ 
		& 2\alpha &  &  &  \\ 
		&  & \ddots &  &  \\ 
		&  &  & \left( K-n-1\right) \alpha &  \\ 
		&  &  &  & \left( K-n\right) \alpha%
	\end{array}
	\right) _{\left( K-n+1\right) \times \left( K-n\right) };
	\setlength{\belowdisplayskip}{3pt}
\end{equation*}%
for $n=M$

\begin{equation*}
	\setlength{\abovedisplayskip}{3pt}
	Q_{M,M}=\left( 
	\begin{array}{ccccc}
		-\left( e_{i}+\mu _{i,0}\right) & e_{i} &  &  &  \\ 
		\lambda _{i} & -\left( \lambda _{i}+e_{i}+\mu _{i,0}\right) & e_{i} &  &  \\ 
		& \ddots & \ddots &  &  \\ 
		&  & \lambda _{i} & -\left( \lambda _{i}+e_{i}+\mu _{i,0}\right) & e_{i} \\ 
		&  &  & \lambda _{i} & -\left( \lambda _{i}+\mu _{i,0}\right)%
	\end{array}
	\right) ,
	\setlength{\belowdisplayskip}{3pt}
\end{equation*}%
whose size is $\left( K-M+1\right) \times \left( K-M+1\right) $, and

\begin{equation*}
	\setlength{\abovedisplayskip}{3pt}
	Q_{M,0}=\left( 
	\begin{array}{cccccccc}
		\mu _{i,0} &  &  &  &  &  &  &  \\ 
		& \mu _{i,0} &  &  &  &  &  &  \\ 
		&  & \ddots &  &  &  &  &  \\ 
		&  &  & \mu _{i,0} &  &  &  &  \\ 
		&  &  &  & \mu _{i,0} & 0 & \cdots & 0%
	\end{array}
	\right) _{\left( K-M+1\right) \times \left( K+1\right) }.
	\setlength{\belowdisplayskip}{3pt}
\end{equation*}%

\section*{Appendix B. Proof of Lemma 1}

\textbf{Proof: }It is obvious that $Q_{M,M}^{[\leq M]}=Q_{M,M}$. We use
the inductive method to prove that for the censoring matrix $Q^{[\leq k]}$,  Equation (2) is true for $0\leq k\leq M-1$. Our proof 
contains 
three steps
as follows:

\emph{Step one.} For $k=M-1$, referring to Section 2 of Li (2010), we get

\begin{eqnarray*}
	Q^{[\leq M-1]} &=&\left( 
	\begin{array}{ccccc}
		Q_{0,0} & Q_{0,1} &  &  &  \\ 
		& Q_{1,1} & Q_{1,2} &  &  \\ 
		&  & \ddots & \ddots &  \\ 
		&  &  & Q_{M-2,M-2} & Q_{M-2,M-1} \\ 
		&  &  &  & Q_{M-1,M-1}%
	\end{array}
	\right) \\
	&&+\left( 
	\begin{array}{c}
		0 \\ 
		0 \\ 
		\vdots \\ 
		0 \\ 
		Q_{M-1,M}%
	\end{array}
	\right) \left( -Q_{M,M}\right) ^{-1}\left( 
	\begin{array}{ccccc}
		Q_{M,0} & 0 & \cdots & 0 & 0%
	\end{array}
	\right) \\
	&=&\left( 
	\begin{array}{ccccc}
		Q_{0,0} & Q_{0,1} &  &  &  \\ 
		& Q_{1,1} & Q_{1,2} &  &  \\ 
		&  & \ddots & \ddots &  \\ 
		&  &  & Q_{M-2,M-2} & Q_{M-2,M-1} \\ 
		Q_{M,0}^{[\leq M-1]} &  &  &  & Q_{M-1,M-1}%
	\end{array}
	\right) ,
\end{eqnarray*}
in which $Q_{M,0}^{[\leq M-1]}=Q_{M-1,M}\left( -Q_{M,M}\right) ^{-1}Q_{M,0}$
. Thus when $k=M-1$, Equation (2) is true.

\emph{Step two.} We assume that when $k=m$ for $2\leq m\leq M-2$, Equation (2) is true, i.e.,

\begin{equation*}
	\setlength{\abovedisplayskip}{3pt}
	Q^{[\leq k]}=Q^{[\leq m]}=\left( 
	\begin{array}{ccccc}
		Q_{0,0} & Q_{0,1} &  &  &  \\ 
		& Q_{1,1} & Q_{1,2} &  &  \\ 
		&  & \ddots & \ddots &  \\ 
		&  &  & Q_{m-1,m-1} & Q_{m-1,m} \\ 
		Q_{m,0}^{[\leq m]} &  &  &  & Q_{m,m}%
	\end{array}
	\right) ,
	\setlength{\belowdisplayskip}{3pt}
\end{equation*}%
in which $Q_{m,0}^{[\leq m]}=\prod\nolimits_{l=m}^{M-1}\left[
Q_{l,l+1}\left( -Q_{l+1,l+1}\right) ^{-1}\right] Q_{M,0}$. Now, we consider
the case with $k=m-1$ for $2\leq m\leq M-2$,%
\begin{eqnarray*}
	Q^{[\leq m-1]} &=&\left( 
	\begin{array}{ccccc}
		Q_{0,0} & Q_{0,1} &  &  &  \\ 
		& Q_{1,1} & Q_{1,2} &  &  \\ 
		&  & \ddots & \ddots &  \\ 
		&  &  & Q_{m-2,m-2} & Q_{m-2,m-1} \\ 
		&  &  &  & Q_{m-1,m-1}%
	\end{array}
	\right) \\
	&&+\left( 
	\begin{array}{c}
		0 \\ 
		0 \\ 
		\vdots \\ 
		0 \\ 
		Q_{m-1,m}%
	\end{array}
	\right) \left( -Q_{m,m}\right) ^{-1}\left( 
	\begin{array}{ccccc}
		Q_{m,0}^{[\leq m]} & 0 & \cdots & 0 & 0%
	\end{array}
	\right) \\
	&=&\left( 
	\begin{array}{ccccc}
		Q_{0,0} & Q_{0,1} &  &  &  \\ 
		& Q_{1,1} & Q_{1,2} &  &  \\ 
		&  & \ddots & \ddots &  \\ 
		&  &  & Q_{m-2,m-2} & Q_{m-2,m-1} \\ 
		Q_{m-1,0}^{[\leq m-1]} &  &  &  & Q_{m-1,m-1}%
	\end{array}
	\right) ,
\end{eqnarray*}
in which $Q_{m,0}^{[\leq m-1]}=Q_{m-1,m}\left( -Q_{m,m}\right)
^{-1}Q_{m,0}^{[\leq m]}$. Since $Q_{m,0}^{[\leq m]}=\prod\nolimits_{l=m}^{M-1}[Q_{l,l+1}\left(  -Q_{l+1,l+1}%
\right)  ^{-1}]Q_{M,0}$, it is easy to check that $Q_{m,0}^{[\leq
	m-1]}=\prod\nolimits_{l=m-1}^{M-1}\left[ Q_{l,l+1}\left( -Q_{l+1,l+1}\right)
^{-1}\right] Q_{M,0}$. Thus when $k=m-1$, Equation (2) is also true.
Therefore, Equation (2) is true for for $1\leq k\leq M-1$.

\emph{Step three.} It is easy to see from Step two that%
\begin{equation*}
	\setlength{\abovedisplayskip}{3pt}
	Q^{[\leq 1]}=\left( 
	\begin{array}{cc}
		Q_{0,0} & Q_{0,1} \\ 
		Q_{1,0}^{[\leq 1]} & Q_{1,1}%
	\end{array}
	\right) ,
	\setlength{\belowdisplayskip}{3pt}
\end{equation*}%
in which $Q_{1,0}^{[\leq 1]}=\prod\nolimits_{l=1}^{M-1}\left[
Q_{l,l+1}\left( -Q_{l+1,l+1}\right) ^{-1}\right] Q_{M,0}$. Then 
\begin{eqnarray*}
	Q^{[\leq 0]} &=&Q_{0,0}+Q_{0,1}(-Q_{1,1})^{-1}Q_{1,0}^{[\leq 1]} \\
	&=&Q_{0,0}+Q_{0,1}(-Q_{1,1})^{-1}\prod\nolimits_{l=1}^{M-1}\left[
	Q_{l,l+1}\left( -Q_{l+1,l+1}\right) ^{-1}\right] Q_{M,0}.
\end{eqnarray*}

Based on the above three steps, Equation (3) is true for $0\leq k\leq M$.
This completes the proof. \textbf{{\rule{0.08in}{0.08in}}}

\section*{Appendix C. Proof of Theorem 1}

\textbf{Proof:} The proof can easily be completed by means of Chapter 2 of %
Li (2010). Here, it is necessary to fix the two special
matrices $R_{U\text{ }}$and $G_{L}$ as follows:

\textit{The R-measures}: From Equations (4) and (5), we obtain that 
$R_{i,j}=Q_{i,j}^{\left[ \leq j\right] }\left( -Q_{j,j}^{[\leq j]}\right)
^{-1}$ for $0\leq i<j\leq M$. It is seen from Equation (2) that $Q_{i,i+1}^{%
	\left[ \leq i+1\right] }>0$, while the element $Q_{i,j}^{\left[ \leq i+1 %
	\right] }$ is zero for $i+2\leq j\leq M$. Thus for $0\leq i<j\leq M$ 
\begin{equation*}
	\setlength{\abovedisplayskip}{3pt}
	R_{i,j}=\left\{ 
	\begin{array}{ll}
		R_{i,i+1}, & j=i+1, \\ 
		0, & i+2\leq j\leq M.%
	\end{array}
	\right.
	\setlength{\belowdisplayskip}{3pt}
\end{equation*}%
Thus we have%
\begin{equation*}
	\setlength{\abovedisplayskip}{3pt}
	R_{U}=\left( 
	\begin{array}{ccccccc}
		0 & R_{0,1} &  &  &  &  &  \\ 
		& 0 & R_{1,2} &  &  &  &  \\ 
		&  & 0 & R_{2,3} &  &  &  \\ 
		&  &  & \ddots & \ddots &  &  \\ 
		&  &  &  & 0 & R_{M-2,M-1} &  \\ 
		&  &  &  &  & 0 & R_{M-1,M} \\ 
		&  &  &  &  &  & 0%
	\end{array}
	\right) .
	\setlength{\belowdisplayskip}{3pt}
\end{equation*}%

\textit{The G-measures}: From Equations (4) and (6), we obtain that 
$G_{i,j}=\left( -Q_{i,i}^{[\leq i]}\right) ^{-1}Q_{i,j}^{\left[ \leq i\right]
},$ $0\leq j<i\leq M$. From Equation (2), we can see $Q_{i,0}^{\left[ \leq
	i \right] }>0$, while the elements $Q_{i,j}^{\left[ \leq i\right] }$ is zero
for $1\leq j\leq i-1$. Thus for $0\leq j<i\leq M$ 
\begin{equation*}
	\setlength{\abovedisplayskip}{3pt}
	G_{i,j}=\left\{ 
	\begin{array}{ll}
		G_{i,0}, & j=0, \\ 
		0, & 1\leq j\leq i-1.%
	\end{array}
	\right.
	\setlength{\belowdisplayskip}{3pt}
\end{equation*}%
Then we get 
\begin{equation*}
	\setlength{\abovedisplayskip}{3pt}
	G_{L}=\left( 
	\begin{array}{ccccccc}
		0 &  &  &  &  &  &  \\ 
		G_{1,0} & 0 &  &  &  &  &  \\ 
		G_{2,0} &  & 0 &  &  &  &  \\ 
		\vdots &  &  & \ddots &  &  &  \\ 
		G_{M-2,0} &  &  &  & 0 &  &  \\ 
		G_{M-1,0} &  &  &  &  & 0 &  \\ 
		G_{M,0} & 0 & 0 & \cdots & 0 & 0 & 0%
	\end{array}
	\right) .
	\setlength{\belowdisplayskip}{3pt}
\end{equation*}%
Based on the above analysis, this completes the proof. \textbf{{\rule%
		{0.08in}{0.08in}}}

\section*{Appendix D. The block elements of Matrix $\mathbf{T}$}

Note that $\overline{w}\left( n_{B}^{\left( 0\right) }\right) =\min \left\{
n_{B}^{\left( 0\right) },r\right\} w$. For $0\leq k\leq \psi -1$ and
Equation (9), we have%
\begin{equation*}
	\setlength{\abovedisplayskip}{3pt}
	T_{kM,kM}=\left( 
	\begin{array}{ccccc}
		-e_{0} & e_{0} &  &  &  \\ 
		& -\left( \overline{w}\left( M\right) +e_{0}\right) & e_{0} &  &  \\ 
		&  & \ddots & \ddots &  \\ 
		&  &  & -\left( \overline{w}\left( \left( \phi -k-1\right) M\right)
		+e_{0}\right) & e_{0} \\ 
		&  &  &  & -\overline{w}\left( \left( \phi -k\right) M\right)%
	\end{array}
	\right) ,
	\setlength{\belowdisplayskip}{3pt}
\end{equation*}
\begin{equation*}
	\setlength{\abovedisplayskip}{3pt}
	T_{kM,kM+1}=\left( 
	\begin{array}{ccccc}
		0 &  &  &  &  \\ 
		\overline{w}\left( M\right) &  &  &  &  \\ 
		& \overline{w}\left( 2M\right) &  &  &  \\ 
		&  & \ddots &  &  \\ 
		&  &  & \overline{w}\left( \left( \phi -k-1\right) M\right) &  \\ 
		&  &  &  & \overline{w}\left( \left( \phi -k\right) M\right)%
	\end{array}
	\right) ;
	\setlength{\belowdisplayskip}{3pt}
\end{equation*}%
For $0\leq k\leq \psi -1$, $1\leq j\leq M-1$

\begin{equation*}
	\setlength{\abovedisplayskip}{3pt}
	T_{kM+j,kM+j}=\left( 
	\begin{array}{cccc}
		-\left( \overline{w}\left( M-j\right) +e_{0}\right) & e_{0} &  &  \\ 
		& \ddots & \ddots &  \\ 
		&  & -\left( \overline{w}\left( \left( \phi -k-1\right) M-j\right)
		+e_{0}\right) & e_{0} \\ 
		&  &  & -\overline{w}\left( \left( \phi -k\right) M-j\right)%
	\end{array}
	\right) ,
	\setlength{\belowdisplayskip}{3pt}
\end{equation*}
\begin{equation*}
	\setlength{\abovedisplayskip}{3pt}
	T_{kM+j,kM+j+1}=\left( 
	\begin{array}{ccccc}
		\overline{w}\left( M-j\right) &  &  &  &  \\ 
		& \ddots &  &  &  \\ 
		&  & \overline{w}\left( \left( \phi -k-1\right) M-j\right) &  &  \\ 
		&  &  & \overline{w}\left( \left( \phi -k\right) M-j\right) & 
	\end{array}
	\right) ;
	\setlength{\belowdisplayskip}{3pt}
\end{equation*}%

\begin{equation*}
	\setlength{\abovedisplayskip}{3pt}
	T_{Z,Z}=\left( 
	\begin{array}{ccccc}
		-\mu _{0} &  &  &  &  \\ 
		& -\mu _{0} &  &  &  \\ 
		&  & \ddots &  &  \\ 
		&  &  & -\mu _{0} &  \\ 
		&  &  &  & -\mu _{0}%
	\end{array}
	\right) ,
	\setlength{\belowdisplayskip}{3pt}
\end{equation*}
\begin{equation*}
	\setlength{\abovedisplayskip}{3pt}
	T_{Z,0}=\left( 
	\begin{array}{cccccccc}
		\mu _{0} &  &  &  &  &  &  &  \\ 
		& \mu _{0} &  &  &  &  &  &  \\ 
		&  & \ddots &  &  &  &  &  \\ 
		&  &  & \mu _{0} &  &  &  &  \\ 
		&  &  &  & \mu _{0} & 0 & \cdots & 0%
	\end{array}
	\right) .
	\setlength{\belowdisplayskip}{3pt}
\end{equation*}%

\section*{Appendix E. The proof of the Theorem 2}

We first classify the nodes in the closed queueing network
as three classes: Node $i$ for $0\leq i\leq N$; Node $i\rightarrow j$ for $%
1\leq i\leq N$ and $j\in \Theta _{i}$; and Node $i\rightarrow 0$ and Node $%
0\rightarrow i$ for $1\leq i\leq N$. Based on this, we prove the
product-form solution from the following three different parts.

\textit{Part one}: Node $i$ for $0\leq i\leq N$. By using the stationary
probabilities of block-structured Markov processes of base $i$ with $1\leq
i\leq N$ given in Section 4.2, we get 
\begin{equation*}
	\setlength{\abovedisplayskip}{3pt}
	H\left( n_{G}^{\left( i\right) },n_{B}^{\left( i\right) }\right) =\pi
	_{G,n_{G}^{\left( i\right) };B,n_{B}^{\left( i\right) }}^{\left( i\right) }.
	\setlength{\belowdisplayskip}{3pt}
\end{equation*}
Similarly, from the stationary probabilities of block-structured Markov
process of the maintenance shop given in Section 4.3, we obtain 
\begin{equation*}
	\setlength{\abovedisplayskip}{3pt}
	H\left( n_{G}^{\left( 0\right) },n_{B}^{\left( 0\right) }\right) =\pi
	_{B,n_{B}^{\left( 0\right) };G,n_{G}^{\left( 0\right) }}^{\left( 0\right) }.
	\setlength{\belowdisplayskip}{3pt}
\end{equation*}
Thus we have 
\begin{equation*}
	\setlength{\abovedisplayskip}{3pt}
	H\left( n_{G}^{\left( i\right) },n_{B}^{\left( i\right) }\right) =\left\{ 
	\begin{array}{ll}
		\pi _{G,n_{G}^{\left( i\right) };B,n_{B}^{\left( i\right) }}^{\left(
			i\right) }, & 1\leq i\leq N,0\leq n_{G}^{\left( i\right) }\leq K,0\leq
		n_{B}^{\left( i\right) }\leq M, \\ 
		\pi _{B,n_{B}^{\left( 0\right) };G,n_{G}^{\left( 0\right) }}^{\left(
			0\right) }, & \text{ }i=0,0\leq n_{B}^{\left( 0\right) }\leq Z,0\leq
		n_{G}^{\left( 0\right) }\leq \phi M.%
	\end{array}%
	\right.
	\setlength{\belowdisplayskip}{3pt}
\end{equation*}%

\textit{Part two}: For Node $i\rightarrow j$ with $1\leq i\leq N$ and $j\in
\Theta _{i}$.

Since the number of servers in Node $i\rightarrow j$ may be infinite,
applying  Gordon and Newell (1967) and Subsection 7.35 in Bolch et al. (2006), the function $H\left( m_{i,j}\right) $ is given by
\begin{equation*}
	\setlength{\abovedisplayskip}{3pt}
	H\left( m_{i,j}\right) =\left( \frac{e_{i,j}}{\mu _{i,j}}\right) ^{m_{i,j}}%
	\frac{1}{m_{i,j}!},\text{ }0\leq m_{i,j}\leq K.
	\setlength{\belowdisplayskip}{3pt}
\end{equation*}%

\textit{Part three}: For Node $i\rightarrow 0$ (resp. Node $0\rightarrow i$)
with $1\leq i\leq N$.

The failed items (resp. the repaired items) are transported on path nodes by
the batch removal (resp. batch redistribution) policy. Refer to Henderson and Taylor (1990), we get 
\begin{equation*}
	\setlength{\abovedisplayskip}{3pt}
	H\left( m_{i,0}\right) =\left( \frac{e_{i,0}}{\mu _{i,0}}\right) ^{m_{i,0}}%
	\frac{1}{m_{i,0}!},\text{ }m_{i,0}=kM\text{ for }0\leq k\leq \phi ,
	\setlength{\belowdisplayskip}{3pt}
\end{equation*}
and%
\begin{equation*}
	\setlength{\abovedisplayskip}{3pt}
	H\left( m_{0,i}\right) =\left( \frac{e_{0,i}}{\mu _{0,i}}\right) ^{m_{0,i}}%
	\frac{1}{\left( m_{0,i}\right) !},\text{ }m_{0,i}=lZ_{i}\text{ for }0\leq
	l\leq \frac{\phi }{\psi }.
	\setlength{\belowdisplayskip}{3pt}
\end{equation*}%

Finally, for the closed queueing network, we can express the product-form
solution in Theorem 2, and the normalization constant is given by 
\begin{equation*}
	\setlength{\abovedisplayskip}{3pt}
	\mathbf{C}=\sum\limits_{\mathbf{n}\in \Omega }\prod\limits_{i=0}^{N}H\left(
	n_{G}^{\left( i\right) },n_{B}^{\left( i\right) }\right) \prod\limits 
	_{\substack{ i=1  \\ j\in \Theta _{i}}}^{N}H\left( m_{i,j}\right)
	\prod\limits_{i=1}^{N}H\left( m_{i,0}\right) H\left( m_{0,i}\right) .
	\setlength{\belowdisplayskip}{3pt}
\end{equation*}
This completes the proof. \textbf{{\rule{0.08in}{0.08in}}}

\bibliographystyle{apalike}
\bibliography{DBSS}

\end{document}